\documentclass[galaxies,article,accept,moreauthors,pdftex]{mdpi} 
\usepackage{xcolor}
\usepackage{mathtools}

\DeclarePairedDelimiter\abs{\lvert}{\rvert}%
\firstpage{1} 
\pubvolume{11}
\issuenum{6}
\articlenumber{110}
\pubyear{2023}
\copyrightyear{2023}
\externaleditor{Academic Editor: Alberto C. Sadun} 
\datereceived{31 July 2023}
\daterevised{18 October 2023}
\dateaccepted{30 October 2023} 
\datepublished{9 November 2023} 
\hreflink{https://doi.org/10.3390/\linebreak galaxies11060110} 

\Title{Simulated Radio and Neutrino Imaging of a Microquasar}

\TitleCitation{Simulated Radio and Neutrino Imaging of a Microquasar}

\Author{{Theodoros Smponias} 
}

\AuthorNames{Theodoros Smponias}

\AuthorCitation{Smponias, T.}

\address[1] {%
 Directorate of Primary Education of the Ionian Islands, 49100 Corfu City, {Corfu}
, Greece; t.smponias@hushmail.com}

\abstract{\textls[-5]{Microquasar stellar systems emit electromagnetic radiation and high-energy particles. Thanks to their location within our own galaxy, they can be observed in high detail. Still, many of their inner workings remain elusive; hence, simulations, as the link between observations and theory, are highly useful. In this paper, both high-energy particle  and synchrotron radio emissions from simulated microquasar jets are calculated using special relativistic imaging. A finite ray speed imaging algorithm is employed  on hydrodynamic simulation data, producing synthetic images seen from a stationary observer. A hydrodynamical model is integrated in the above emission models. Synthetic spectra and maps are then produced that can be compared to observations from detector arrays. As an application, the model synthetically observes microquasars during an episodic ejection at two different spatio-temporal scales: one on the neutrino emission region scale and the other on the synchrotron radio emission scale. The results are compared to the sensitivity of existing detectors.} 
}


\begin{document}

\section{Introduction} 
\label{intro}

Microquasars (MQs) include a binary stellar system, with a main sequence star orbiting a collapsed stellar remnant \cite{Mirabel99}. Material from the star accretes onto a compact object, resulting in the production of a pair of relativistic ejections moving in opposite directions largely perpendicular to the binary orbital plane. These ejecta form  jets, which emit anywhere between {radio waves} 
and very-high-energy (VHE) $\gamma$ rays and neutrinos \cite{Romero2003,Bednarek2005,Bosch_Ramon_2007,Reynoso2008,Reynoso2009,Christiansen2013high,Zhang2010,Reynoso2019,Reynoso_Romero_2021,Romero_22mq,Romero_23_twin_bh,GRS_april23}.  
As shown in \cite{Romero2003}, the apparent superluminal motion in MQs suggests that jets comprise bulk hadron flows. The assumption of equipartition \cite{Reynoso2009} leads to high magnetic field estimates, especially for inner jets \cite{Koessl1990}. The latter, together with the fluid approximation for the jet matter due to the presence of tangled magnetic fields \cite{Rieger2006,Rieger2019}, lays the foundation for the jet  magnetohydrodynamic (MHD) approximation. A toroidal magnetic field may contribute to jet collimation along its path \cite{Koessl1990,Singh2019MHD}, while confinement from surrounding winds is also a possibility \cite{Hughes1991,Reynoso2009}.

  
A turbulent fluid jet region can give rise to a variety of signals, from radio waves to \linebreak $\gamma$ rays. Furthermore, cascades of high-energy particles in the jets produce different particle populations linked by the transport phenomenon. The emergence of neutrinos that then leave the system can be detected by modern arrays. 

In this work, the production of radio synchrotron emission and  very-high-energy (VHE; up to hundreds of TeV) neutrinos from generic MQ jets is studied using the method of dynamic and radiative relativistic MHD simulations, where the model space is divided into computational cells. The effects of wind from a companion star are also included.

Adopting the one-zone (homogeneous) emission model in each eligible hydrocode cell (in this work, hydrocode refers to the PLUTO program \cite{Mignone2007}), the solution of successive transport equations connecting particle distributions in a cascade provides the emitted neutrino intensity as a function of local dynamic and radiative jet parameters. Particle cascade calculations, acting on hydrocode data outputs, employ the results from Monte Carlo simulations \cite{Kelner2006, Lipari2007}. This way, a cell's physical parameters are directly connected to its particle emission. The cascade timescale is assumed to be much smaller than the dynamic timescale of the modeled system. Repeating the above calculations over a number of different energies provides a neutrino energy spectrum at each jet point. Line-of-sight integration follows, leading to the production of synthetic neutrino images and  model system spectra. 

Charge neutrality is assumed in the jets, coupling the bulk flow proton and electron distributions dynamically through a relativistic MHD simulation. The latter provides  the bulk proton density and also the magnetic field in each cell. High-energy hadron and lepton populations, obtained through shock front acceleration in the jets using the one-zone model in each computational cell, lead to synchrotron emission \cite{Reynoso2009} taking into account local magnetic fields. Here, we focus on the radio synchrotron band, where the spectrum is flat or inverted and the emission from protons is essentially negligible (see Figure 5 in~\cite{Reynoso2009}). As shown in the {Appendix} 
 \ref{energetics}  (Appendix \ref{sizing}), the inner jet region of $\simeq$10$^{-3}$ pc is modeled here (corresponding to roughly 10 mas, at a distance to Earth of 5 kpc), where a flat radio spectrum typically occurs. 

Lepton and hadron contributions that are comparable to radio synchrotron only occur when much more energy is assigned to hadrons than to leptons. As an approximation, only synchrotron emission from accelerated jet electrons is modeled here using the formalism in~\cite{Pacholczyk_1970}. Self-absorption is also included. 


Consequently, local MHD jet properties, as provided by the MHD model, allow the calculation of local radio synchrotron emission and absorption coefficients. Down the pipeline, this may lead to synthetic radio maps of the model system. When combined, the study of both synchrotron and neutrino emission from the jets may offer an improved understanding of high-energy processes in the system.

In the current work, the finite nature of the speed of light is  taken into account when viewing the model jet, which constitutes an improvement over previous work \cite{Smponias_2021}. Imaging a relativistically moving macroscopical object opens the window to a rather unexpected and even strange world of peculiarities. The basic mandates of Special Relativity, regarding length contraction and time dilation, constitute the mere beginning in the quest for comprehension of the actual appearance of a fast-moving object \cite{Lampa24, Terrell59, Penrose59, Weisskopf1960, SV65}. An observer will see the object view affected by a number of relativistic distortions \cite{Weiskopf01, Deissler2005, Weiskopf10}.

Electromagnetic emission transformation from the jet frame of reference to the Earth frame requires performing the Lorentz/Poincar\'e transform \cite{Cawthorne1991, Hughes1991, WKR99, Weiskopf01}. Applying the latter transform for imaging purposes aims to reconstruct what the observer will actually see. Relevant to this point, \cite{Klauber2008} argues about the important difference between vision and measurements in Special Relativity, presenting that difference in a geometrical~manner.

Radiation emitted from a jet is therefore subject to relativistic effects \cite{RL79, Hughes1991}, including  time dilation, relativistic aberration, and frequency shift,  collectively leading to what is known as Doppler boosting and beaming \cite{WKR99, Weiskopf01, Jarabo2015}. Aberration causes the fast-moving object to actually appear rotated to a stationary observer  \cite{Terrell59, Penrose59, Hickey79, Weiskopf01}, a phenomenon sometimes called the Terell--Penrose rotation.

Ray-tracing methods provide excellent-quality  relativistic images, despite lacking in terms of efficiency compared to such techniques as polygon rendering \cite{Weiskopf10}. In this work, a relativistic imaging method is employed, whereby time-resolved hydrocode data are  crossed by rays called lines of sight (LOSs), either focused or parallel to each other. Each ray is then subject to beaming effects, which means Doppler boosting for E/M radiation, depending on the ray and the local velocity relative orientation.


\textls[-15]{Furthermore, the aforementioned high-energy proton distribution is Lorentz-transformed} to a stationary frame, exhibiting dependence on speed and orientation, favoring particle emissions along the direction of the local velocity \cite{TR11}.  The resulting expression demonstrates relativistic boosting properties in a manner broadly comparable to the Doppler boosting of E/M radiation. A cascade of particle distributions then emerges, affected by local MHD properties, eventually leading to neutrino emission, which then escapes the system. Subsequently, with neutrinos, imaging ``rays'' move at the finite speed of light.

Simulating the above processes may help clarify the inner workings of the jets and their environs, leading to a more accurate description of the system of interest. 

In the remainder of this paper, the theoretical background is presented for both radio synchrotron and neutrino particle emissions (Section \ref{background}). Section \ref{software} briefly describes the employed software pipeline. The model setup is presented next (Section \ref{model_setup}), including various model parameters. In Section \ref{results}, the results are presented and discussed. Finally, in Section \ref{conclusions}, useful conclusions are drawn from the current work and possible future applications are proposed.

\section{Theoretical Background}
\label{background}













A conceptual link between electromagnetic and particle emissions from the jets may be formed. Proposed neutrino emission from jets favors the presence of high-energy protons and electrons, triggering particle cascades that lead to neutrinos. Jets are therefore acceleration sites, possibly at shock fronts inside them. High-energy electrons lay the foundation for synchrotron radio wave emission, as well as emission in other parts of the electromagnetic spectrum. Shock front acceleration therefore energizes the jets, leading to both particle and radiation production thereafter.

\subsection{Radio Synchrotron Emission and Self-Absorption}


Adopting high-energy electron and proton acceleration at shock fronts \cite{Pacholczyk_1970}, a power-law high energy distribution is assumed for each of the aforementioned particle species. Electrons are discussed in this subsection; protons are discussed in the following one. For~electrons, 
\begin{equation}
N_{(E)} = N_{\rm{0}} E^{- \gamma_{c}} = K_{\rm{pl}} \rho_{\rm{4d}} E^{- \gamma_{c}},
\label{electron_power_law}
\end{equation}
where  $K_{\rm{pl}}$ is a power law constant for high-energy jet electrons. $K_{\rm{pl}} \rho_{\rm{4d}}= N_{\rm{0}}$; therefore, the reference electron density N$_{\rm{0}}$ is taken as proportional to the local bulk flow proton density $\rho_{\rm{4d}}$ based on charge neutrality. $\rho_{\rm{4d}}$ is provided for spatiotemporal points by MHD code. Thus, a link is established between PLUTO's hydrodynamic quantities and electron populations in the jet.

For accelerated protons and electrons, a high-energy cutoff is adopted according to the Hillas criterion \cite{Lobato_Coelho_2017}.  $\mathrm{E}_{\rm{max}} = \mathrm{Z q B} \mathrm{R}_{\rm{s}}$ is the maximum energy {E$_{\rm{max}}$} 
 achieved through acceleration for a particle of charge q and atomic number {Z} within a magnetic field B in an accelerator of size $\mathrm{R}_{\rm{s}}$. 

For the radio-scale model, the computational cell size of 10$^{13}$ cm is selected in order to fit the radio-resolved portion of the jet into the model space. Furthermore, the magnetic field is set at a value less than equipartition, at 10 G. The above equation will then result, for electrons, in a maximum energy  on the TeV scale, well above the radio band which is of interest in this subsection. Furthermore, the energy cutoff from dominant cooling mechanisms \cite{Reynoso2009} for radio-band synchrotron from a microquasar lies above radio wave energies. Thus, no cutoff is needed for electrons here. 

\textls[-25]{On the other hand, for high-energy protons, a cutoff of E$_{\rm{max}}$ = 10$^{6}$ GeV is employed~\mbox{\cite{Reynoso2009,Smponias_2021}.}}

Synchrotron emission and absorption coefficients are then calculated, following \cite{Pacholczyk_1970}:\vspace{-12pt}

\begin{equation}
\epsilon_{\nu}=f_{\rm{pp}(\rm{4d})} K_{\rm{pl}} c_{5(\gamma_{c})} \rho_{\rm{4d}} (\overrightarrow{B_{\rm{4d}}} \sin(\widehat{\overrightarrow{\rm{LOS}},\overrightarrow{B_{\rm{4d}} }} ) )^{  \frac{(\gamma_{c}+1) }{2}  } \left( \frac{f_{\rm{obs} }}{2 D_{\rm{4D}_{local} } c_{1} } \right)^{ \frac{(1-\gamma_{c})}{2} }   D_{\rm{4D}_{local} }^{2}
\end{equation} 
\begin{equation}
k_{\nu}=f_{\rm{pp}(\rm{4d})} K_{\rm{pl}} c_{6(\gamma_{c})} \rho_{\rm{4d}} (\overrightarrow{B_{\rm{4d}}} \sin(\widehat{\overrightarrow{\rm{LOS}},\overrightarrow{B_{\rm{4d}}}} ) )^{\frac{(\gamma_{c}+2)}{2}} \left(\frac{f_{\rm{obs} } }{2 D_{\rm{4D}_{local} } c_{1} } \right)^{\frac{(-4-\gamma_{c})}{2}} D_{\rm{4D}_{local} }^{2},
\end{equation}
where f$_{\rm{pp(4d)}}$ is a proton profile function, with a threshold velocity u$_{\mathrm{th}}$ of {0.1 c} 
 (f$_{\mathrm{pp(4d)}}$ = 1.0 if $\abs{(u/u_{\mathrm{th}})}$ > 1, and f$_{\mathrm{pp(4d)}}$ = $\abs{(u/u_{\mathrm{th}})}^{z_{\mathrm{fpp}}}$ if $\abs{(u/u_{\mathrm{th}})}$ < 1); $f_{\rm{obs}}$ is the observing frequency; and $\frac{f_{\rm{obs}}}{D_{\rm{4D}_{local} } }$ is the calculation frequency (blue-shifted or red-shifted according to the relative orientation between the local LOS and local velocity); $\overrightarrow{B_{\rm{4d}}}$ is the local magnetic field vector; $\overrightarrow{\rm{LOS}}$ is the line of sight (LOS) vector at an angle ($ \widehat{ \overrightarrow{\rm{LOS}},\overrightarrow{B_{\rm{4d} } }  } $) to the magnetic field spatial vector; $D_{\rm{4D_{local}}}$ is the local Doppler factor of E/M emission from a computational cell, calculated as $D=\frac{\sqrt{1-u^{2}}}{(1-u* \cos(\rm{losu}))}$, 
$\rm{losu}=(\widehat{\overrightarrow{\rm{LOS}},\overrightarrow{u}})$ being the angle between the LOS and the local velocity u, 0 $\leq$ u $\leq$ 1 (Appendix \ref{Appendix_coslosu_calc}); and $\gamma_{c}$ is the electron power law index, taken as 2 in the current work, equal---for simplicity---to the proton power law index. According to~\cite{Pacholczyk_1970}, the quantities $c_{i}$, $i$ = [1,~6], are
$c_{1}$ = 6.27 $\times$ 10$^{18}$, $c_{2}$ = 2.37 $\times$ 10$^{-3}$, \linebreak $c_{3}$ = 1.87 $\times$ 10$^{-23}$, \mbox{$c_{4}$ = 4.2 $\times$ 10$^{7}$}
\begin{equation}
c_{5}=0.25 c_{3} \Gamma \left(\frac{3  \gamma_{c}-1}{12} \right) \Gamma \left(\frac{3 \gamma_{c}+7}{12} \right) \left( \frac{\gamma_{c} + \frac{7}{3} }{\gamma_{c}+1} \right)
\end{equation}
\begin{equation}
c_{6}=\frac{1}{32} \left(\frac{3 \times 10^{10}}{c_{1} } \right)^{2} c_{3} \left(\gamma_{c}+\frac{10}{3} \right) \Gamma \left(\frac{3  \gamma_{c}+2}{12} \right) \Gamma \left(\frac{3  \gamma_{c}+10}{12} \right),
\end{equation}
where $\Gamma$ is the Gamma function.

The above coefficients are computed at each point (computational cell) of the model jet system. The angle ($\widehat{\overrightarrow{\rm{LOS}},\overrightarrow{B_{\rm{4d} }}}$) between the LOS and the local magnetic field B is calculated at each jet point, Appendix \ref{Appendix_coslosb_calc}. 

In the above, only two Doppler factors and a frequency shift factor are included. The reason for this is to allow detection of the receding blob in the synthetic radio images. Additional Doppler factors can be employed, see Appendix \ref{Dboosting}.

\subsection{Nonthermal Proton Density}

Neutrino emission from model jets comes from proton--proton interactions between a distribution of high-energy (non-thermal) protons and bulk flow protons \cite{Romero2003,Reynoso2008,Reynoso2009,Reynoso2019,Kelner2006,Lipari2007,Smponias_2021}. Neutrino emission also comes from $\mathrm{p}\gamma$ interactions. For the energy range employed here, from 1 GeV to 5 $\times$ 10$^{5}$ GeV, the pp contribution is higher than the $\mathrm{p}\gamma$ one. According to Figure 7 in \cite{Reynoso2009}, at 1 GeV, the difference is several orders of magnitude in favor of pp, while even at 10$^6$ GeV, the pp contribution remains higher. Therefore, over the energy range of interest, the $\mathrm{p}\gamma$ contribution may, in the first approximation, be left out. Nevertheless, including the $\mathrm{p}\gamma$ contribution is a priority in future use of the model.

Some thermal protons gain energy at shock fronts within the jet according to the first-order Fermi acceleration mechanism within a time frame of \cite{Begelman1980,Rieger2006,Rieger2019}:

\begin{equation}
t^{-1}_{\mathrm{acc}} \simeq \eta \frac{c e B}{E_{\rm{p}}},
\end{equation}
where $B$ is the magnetic field,  $E_{\rm{p}}$ is the non-thermal proton energy, $e$ is the particle charge, and $c$ is the speed of light. $\eta$ is an acceleration efficiency parameter near the base of the jet for the process of efficient particle acceleration in moderately relativistic shocks  \cite{Begelman1980}. In general, $\eta$ depends on the shock velocity and on the local diffusion coefficient \cite{Romero_Muller_Roth_2018_AA}. As an approximation, $\eta=0.1$ \cite{Begelman1980}. 

A power law energy E distribution is employed for non-thermal protons, \linebreak $N_{\rm{p}} = KN_{\rm{p(0)}} E^{-\alpha}$ \cite{Hughes1991}. \textls[-15]{The constant K provides the ratio between the number densities of hot and cold protons, and is much smaller than unity ($N_{\rm{p}}$ is the hot proton number density and $N_{\rm{p(0)}}$ is the cold proton number density). In the model, the proton spectral index in the local jet matter frame, $\alpha$ $\approx$~2  \cite{Reynoso2008}. In general, $\alpha$ is a parameter of the model \cite{Reynoso2019} that depends on the non-thermal proton distribution profile, and may be changed in future work.}

In this work, transport equations are resolved for pions and neutrinos, while a power law is used for hot protons. Nevertheless, a transport equation may also be used in order to find the hot proton distribution \cite{Reynoso2008}. 




In this work, the high-energy proton distribution is considered isotropic in the jet frame. {\textls[-5]{The hypothetical anisotropy of the hot proton distribution can be reflected in the}}
neutrino distribution \cite{Derishev06}, potentially introducing anisotropy in the jet frame's particle emission  field.
 Assuming that $\mathrm{l}_{\rm{sc}} < \mathrm{l}_{\rm{r}}$  at each jet point, where l$_{\rm{sc}}$ is the scattering length and l$_{\rm{r}}$ the radiative length. The above approximation is based on the need to preserve, after every bounce, at least some proton energy \cite{Rieger2019}. The scattering length l$_{\rm{sc}}$ is less than the radiative length l$_{\rm{r}}$, otherwise the proton would not have any energy left after the bounce, negating the acceleration process.




\subsection{Neutrino Emissivity}
\label{neutrino_emissivity}

For each computational cell, a high-energy proton distribution is transformed
 \cite{TR11} from the cell's frame to our frame, using the angle of local velocity to the LOS crossing that cell. In each cell, a local particle cascade emerges. From protons to pions and then to neutrinos, successive particle populations are linked by transport equations. {As an approximation, the branch leading to muonic neutrinos is not included here.} 
 In each cell,  
transport equations are resolved along the particle cascade. Starting from a power law high-energy proton distribution, we obtain a a pion distribution and then a neutrino population, which then leave the system \cite{Kelner2006,Lipari2007,Reynoso2008,Reynoso2009}. The resulting neutrino population is
\begin{eqnarray}
Q_{\pi \rightarrow \nu}(\mathrm{E}) = \int \limits_{E}^{E_{\mathrm{max}}} dE_{\pi} t^{-1}_{\pi} 
(E_{\pi}) N_{\pi}(E_{\pi}) \frac{\Theta (1-r_{\pi}-x)} {E_{\pi}(1-r_{\pi})}  \, ,
\label{Neut-Emiss}
\end{eqnarray}
where E is the neutrino energy, E$_{\pi}$ is the pion energy, N$_{\pi}$ is the pion number density
, $x=\mathrm{E}/\mathrm{E}_{\pi}$, r$_{\pi}$ = (m$_{\mu}$/m$_{\pi}$)$^{2}$ (m$_{\mu}$ and m$_{\pi}$ are the muon and pion masses, respectively) and $t_{\pi}$ is the pion decay timescale. $\Theta$($\chi$) 
is the theta function ~\cite{Reynoso2009,SK15}. As shown in \cite{Reynoso2008},  \linebreak E$_{\rm{max}}$ = 10$^{6}$ GeV. The calculation leading to the above result can be found in \cite{Smponias_2021} (see also the Appendix~\ref{neutrinos}).
Neutrino emission is calculated for each eligible hydrocode cell. The imaging process may incorporate either parallel LOSs or a focused beam, where each LOS follows a slightly different path to a common focal point \cite{RLOS}. A synthetic image of the model system is thus produced. 

\textls[-15]{Muon neutrinos are not included in the first approximation, since their anticipated contribution is perhaps an order of magnitude lower. Their inclusion is deferred to future~work.  }

\section{Computer Programs Used}
\label{software}

\begin{figure}[H]
\includegraphics[width=10.5 cm]{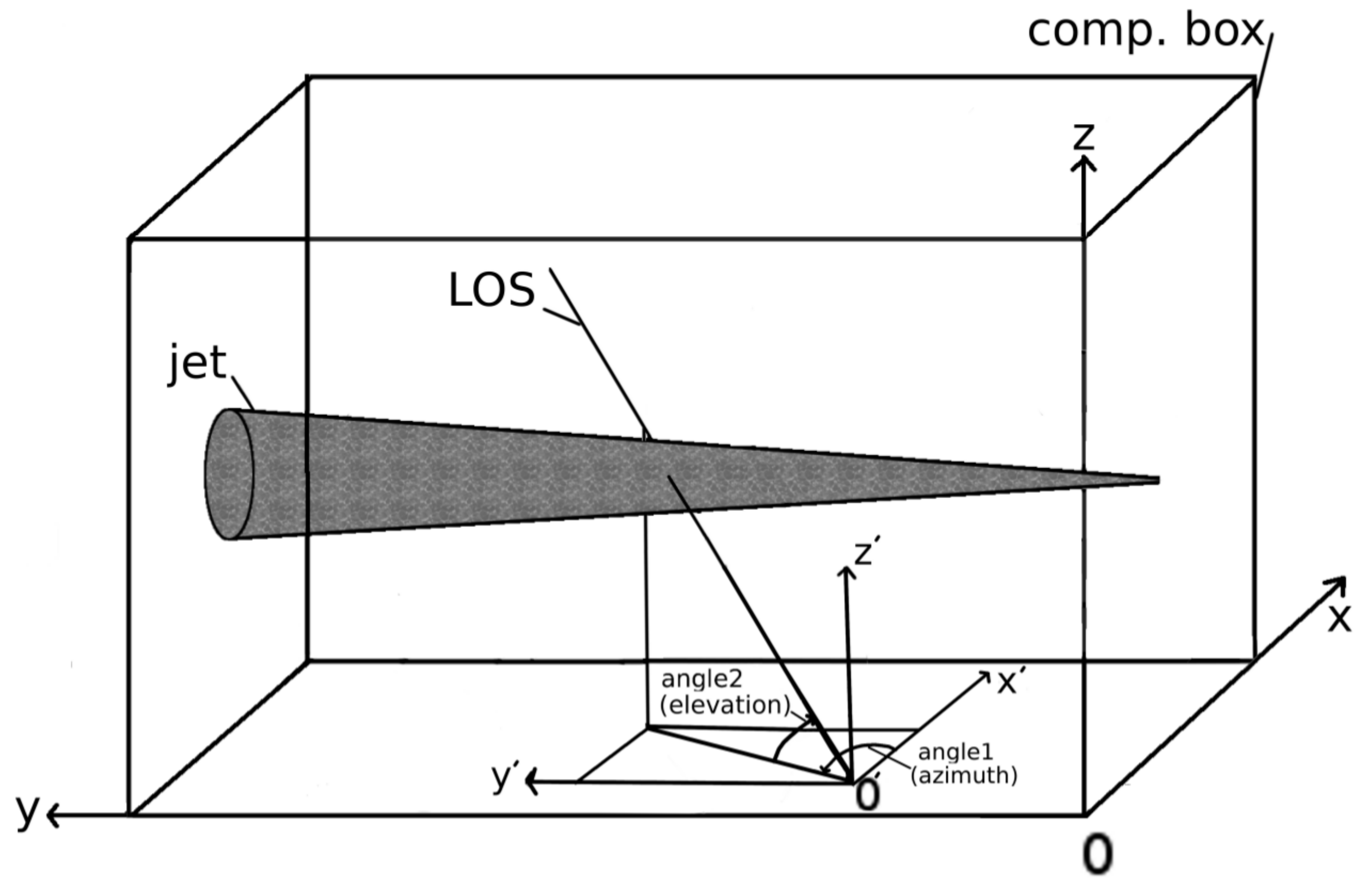}
\caption{{Three-dimensional schematic view} 
 of rlos applied to a model astrophysical jet for a parallel ray setup. The imaging side of the computational box is the yz plane located on the side of the box that appears closer to the reader. Lying on the yz plane, O$^{\prime}$ is the point of origin of a random LOS with its own dashed coordinate system x$^{\prime}$y$^{\prime}$z$^{\prime}$. Alternatively, the imaging plane may also lie on the xz side of the box.}
\label{3Dgeometry}
\end{figure}

\subsection{RLOS}
\label{3dimaging}

rlos \cite{RLOS} is an evolution of classical imaging code used in earlier works \cite{SK11, SK14, KS18}. A ray, or an LOS, emanates from each pixel of the imaging side of the Cartesian 3D computational domain, Figure \ref{3Dgeometry}, or from a focal point, aiming at a point on a fiducial screen, Figure  \ref{focused_beam_imaging_geometry} (Appendix \ref{appc}). Either way, there is an imaging plane. Along the LOS, the   radiative transfer equation is solved in each cell using local emission and absorption coefficients. Depending on the modeled situation, the coefficients may either be directly calculated or outsourced to another program.


rlos is organized into two outer spatial loops running over the imaging plane and an inner one-dimensional spatial loop, advancing in pairs of steps, one for each direction angle (azimuth and elevation), running over the length of an LOS (Figure \ref{flow_diagram1}). At the innermost point of the nested loop lies a conditional temporal loop, running over the hydro data time span (see also the Appendix~\ref{appc}, Appendix \ref{back_in_time}). Since the emission coefficients' calculation load is global, it is performed, where feasible, before the loops in array-oriented operations in order to improve performance.

Lines-of-sight are drawn starting from a focal point (focused beam) or from a pixel of the yz or xz side (parallel rays) of the domain, Figure \ref{3Dgeometry} (see also Figures \ref{losdraw} and \ref{focused_beam_imaging_geometry} in the Appendix~\ref{appc}). Tracing their way along the given direction, they reach a length of $\sqrt{(x_{\rm{max}}^{2}+y_{\rm{max}}^{2}+z_{\rm{max}}^{2})}$, where $x_{\rm{max}}, y_{\rm{max}}, z_{\rm{max}}$ are the cell dimensions  of the computational domain. In practice, on reaching the ends of the domain, an LOS calculation halts, and some LOSs may thereby end up being shorter than others. The above process is repeated within a 2D loop running over the imaging plane, with each LOS corresponding to a single pixel of the final synthetic image. As an approximation, no sideways scattering is considered  along an LOS. 

A model astrophysical system geometry may be directly inserted into rlos. As an example, a ``conical'' jet setup \cite{HJ88} is available to the user. Alternatively, data output from a hydrocode may be employed, as in the current paper, using PLUTO \cite{Mignone2007}.

\subsection{The PLUTO Hydrocode}

PLUTO \cite{Mignone2007} is an open-source, 2D/3D modular hydrocode---a finite-volume/difference shock-capturing program---meant to integrate a set of (time-dependent) conservation laws. Initial and boundary conditions are conveniently assigned through an equivalent set of primitive variables. The relevant systems of equations may include hydrodynamics (HD), magnetohydrodynamics (MHD), and their special-relativistic counterparts, RHD and RMHD, respectively, in either two or three spatial dimensions. The solution of conservation laws is produced through discretization on a structured mesh---a logically rectangular grid surrounded by a boundary with additional ghost cells---in order to implement boundary conditions. The grid may either be static or adaptive, and various coordinate systems are available. The program may run efficiently in parallel on various platforms.

\subsection{Nemiss}
\label{nemiss}

Nemiss \cite{nemiss,Smponias_2021} calculates neutrino emission and spectra from the output of PLUTO hydrocode. It stands between PLUTO and rlos, taking the burden of global particle cascade calculations off the shoulders of rlos, helping form the  PLUTO--nemiss--rlos pipeline. Synthetic neutrino images are produced taking into account the finite speed of emitted neutrinos. Doppler boosting and frequency shifts are switched off in rlos when imaging with neutrinos.

\subsection{Software Information}
\label{pipeline}
Intensity plots of the jet pair  were created using Veusz (\url{https://github.com/veusz}), a software for plotting data written by Jeremy Sanders and contributors and distributed under the GNU/GPL license. rlos and nemiss \cite{nemiss}, written by the author, are available under the lGPL license. PLUTO was written by Andrea Mignone and collaborators, and is available under GNU/GPL.

\section{Model Setup}
\label{model_setup}
 
The MQ system is represented on two different scales: one for modeling neutrino emission and another for radio emission. On the smaller scale, meant for neutrinos, an accretion disk is assumed around the collapsed object \cite{Fabrika2004}, while the companion star itself lies outside the model space. A continuous ejection, representing the beginning of a new blob, is employed. 

On the other hand, on the radio emission scale, both participants of the binary system lie within the same computational cell, while a sequence of plasmoids is  employed.

In both cases, twin jets emerge from near the compacted star, with a collimating toroidal magnetic field component. The field is set higher at the neutrino emission scale.

\subsection{Radio Synchrotron Emission Model}

A twin model jet system is synthetically observed using radio synchrotron emission at a ``typical'' radio frequency of 8 GHz. A 3D homogeneous Cartesian coordinate system is employed. Jets enter the grid at a generic speed of 0.8c (a speed attributed, for example, to the jets of GRS1915+105), emanating from a central location in the grid.



The jet is synthetically observed using synchrotron emission and self-absorption \cite{Pacholczyk_1970}. In each computational cell, the angle between the local magnetic field and the LOS is calculated, leading to determination of the cell's emission and absorption coefficients. Likewise, the local angle between the LOS and velocity enables calculation of each cell's Doppler boosting. 

A sequence of twin relativistic blobs is ejected from the jet base, moving in opposite directions at an angle to the observer. One jet is approaching, the other is receding. The model jet is made of a series of such plasmoids ejected for 30 hydrocode time units every 100~time units. The latter choice of the   plasmoid size is aimed to produce typical intermittent jets made from a series of blobs. Two separate hydrocode runs were performed, one with a heavier jet and another with a lighter jet. 

A special note should be made concerning the employed model jet power. The heavier radio jet in the model is very strong, at about 10$^{43}$ ergs$^{-1}$, whereas the lighter one stands at around 10$^{41}$ ergs$^{-1}$. These values are higher than the Eddington luminosity for an MQ system, especially the heavier jet model. The reason for this is that limited computational resources and the use of a homogeneous grid led to a low model resolution being employed. The latter defines the minimum nozzle diameter, which is much larger than the real jet. Thus, in order to keep jet densities realistic, the jet powers tend to be higher than normal. Details of the runs are shown in Table~\ref{Table-for-run-data}.

 A toroidal magnetic field of 10 G is introduced, resulting in a magnetic energy density below the kinetic energy density (see the {Appendix}
~\ref{appb}). For simplicity, the magnetic field in this run was initially wrapped around the y-axis.  

At regular time intervals, a model system instance was saved to disk. Then, synchrotron emission $\epsilon_{\nu}$ and absorption $\kappa_{\nu}$ coefficients are calculated at each spatial point for a given set of observation angles   within rlos. A series of four-dimensional arrays is thus obtained, through which lines-of-sight travel. Rays start at the moment of observation and move backwards in both space and time in order to meet blobs at an earlier instant. 

The angle to the y axis was 55 degrees in order to enhance the effect of the apparent acceleration of approaching plasmoids in the synthetic radio images. The finite ray speed focused beam imaging method was employed in all radio imaging runs. As a test, an RMHD run was also performed with an angle to the LOS of 55 degrees in order to better visualize apparent superluminal motion. A simplified stellar wind construct was employed; its density was inversely proportional to the distance from the companion star \cite{SK11}.

\begin{table}[H]
\caption{ {Three different imaging} 
 runs based on three separate underlying hydrocode runs.}
\label{Table-for-run-data}
	\begin{adjustwidth}{-\extralength}{0cm}
		\newcolumntype{C}{>{\centering\arraybackslash}X}
		\begin{tabularx}{\fulllength}{LlllL}
			\toprule
			
{\textbf{Model}} & {\textbf{Radio Heavy}} & {\textbf{Radio Light}} & {\boldmath{$\nu$}\textbf{-Scale}} & { \textbf{Comments}} \\
\midrule
{$l_{\mathrm{\rm{cell}}}$ ($\times 10 ^{10} $ cm)} & 2.0 $\times~10 ^{3}$ & 2.0~$\times~10 ^{3}$ & 2.0 & PLUTO cell \\
{$\rho_{\rm{jet}}$ (cm$^{-3}$)} & $1.0 \times 10^{12}$ & $1.0 \times 10^{10}$  & $1.0 \times 10^{11}$  & Jet matter density \\
{ $\rho_{{\rm{w}}}$ (cm$^{-3}$)} & $1.0 \times 10^{13}$ & $1.0 \times 10^{11}$ & $1.0 \times 10^{13}$ & Max wind density \\
{ time unit (s)} & $1.0 \times 10^{3}$ & $1.0 \times 10^{3}$ & $1.0$ & Model time scale (model s) \\
ine
{ $\rho_{\rm{dw}}$ (cm$^{-3}$)} & - &  - & $2.0 \times 10^{13}$ & Max disk wind density  \\
{$t^{\rm{max}}_{\rm{run}}$ (s)} & 242~$\times~10^{3}$  & 242~$\times~10^{3}$ & 204 & Model run time \\
{Method} & P. L.~&  P. L.~& P. L.~& Piecewise linear \\
{Integrator} & M. H.~& M. H.~& M. H.~& MUSCL-Hancock \\
{EOS} & Ideal &  Ideal & Ideal & Equation of state\\
{physics} & RelMHD &  RelMHD & RelMHD & PLUTO setup\\
ine
{B field (G)} & $10$  & $10$   & $1.0 \times 10^{4}$  &  Initial toroidal magnetic field \\
{BinSep (cm)} & subcell  & subcell   & $4.0 \times 10^{12}$  &  Binary separation \\
{ $M_{\rm{BH}}/M_{\odot}$} & - & 5--20 & 3--10 &  VE compact star mass \\
{ $M_{\star}/M_{\odot}$} & - & 10--30 & 10--30 &  Companion mass \\
{ $\beta = v_{\rm{0}}/c$ } & 0.8 & 0.8 & 0.8 & Initial jet speed \\
{ $L^{\rm{p}}_{\rm{k}}$ } & $  10^{44}  $ & $ 10^{42}$ & $2 \times 10^{38}$ &  Jet kinetic luminosity \\
{ Jet type} & int. & int. & cont. &  intermittent or continuous  \\
{ $L^{\rm{p}}_{\rm{k(av)}}$} & $ \simeq$10$^{43}$ & $ \simeq$10$^{41}$ & $2 \times 10^{38}$ &  Average Jet kinetic luminosity \\
{Grid resolution} &  60 $\times$  100 $\times$ 50 &  60 $\times$ 100 $\times$ 50 & 60 $\times$ 100 $\times$ 50 &  PLUTO grid size (xyz) \\
ine
{Imaging method} &  FB &  FB & FB & Focused beam \\
{Time delay} &  on &  on & on &  Normal ray speed\\
{Imaging plane} & YZ & YZ & YZ & Fiducial screen parallel to YZ \\
{Emission} & radio-sync & radio-sync & neutrinos & Synthetic emission type \\
{Code used} & PLUTO-rlos & PLUTO-rlos &  PLUTO-nemiss-rlos & pipeline portion employed \\
			\bottomrule
		\end{tabularx}
	\end{adjustwidth}
\end{table}

\subsection{Neutrino Emission}

In the neutrino-scale model scenario, twin hadronic jets were simulated with PLUTO code \cite{Smponias_2021}. Jets are viewed from the side, while the finite-ray speed imaging mode of rlos was  employed. 
 
For the given model view orientation, neutrino emission was calculated at each eligible spacetime point of the computational grid using nemiss \cite{nemiss}. The neutrino emissivity   is separately calculated in individual computational cells using the angle (los,u) formed between the LOS and the local velocity. The parameters used in this scenario are in Table~\ref{Table-for-run-data}. The binary companion now resides outside the computational grid at the position (400, 0, 400),  a binary separation distance of 4 {$\times$} 
 10$^{12}$ cm, affecting the model with its stellar wind \cite{Smponias_2021}, which falls off as 1/r$^{2}$ away from the star,  a typical setup for  stellar wind. The accretion disk was approximately simulated  as  disk-shaped, and a basic accretion disk wind construct, falling off as 1/y away from the disk, was also included. 

The jet being continuous is a feature compatible with the radio-scale model, since at the $\nu$ scale, the time unit of the model is 1000 times less than in the radio scale. The beginning of the injection process of a single radio plasmoid was conceptually modeled in the neutrino-scale scenario. 
 
Furthermore, rlos \cite{RLOS} was employed, which reads the combined results of PLUTO and nemiss, producing synthetic neutrino images of the model system using the focused beam geometry setup. 

At the $\nu$ scale, the relativistic transformation of the hot proton distribution in \cite{TR11} was employed. The magnetic field is toroidal and was adjusted for equipartition, \linebreak \mbox{$B_{\vec{r}z} =\sqrt{8 \pi \rho_{\vec{r} z} }$~\cite{Rieger2006,Rieger2019}}  ({Appendix}
~\ref{appb}). The above choice of magnetic field is an approximation suitable for the inner jet region, with strong fields threading the jets. External magnetic fields can be taken as smaller \cite{Kolo2017}; therefore, as a first-order approximation, they are omitted from the model's surrounding winds. 






The neutrino emission in each computational cell was calculated using the formalism presented earlier in this paper (Section \ref{neutrino_emissivity}). 
This method, albeit costly, allows one to obtain separate neutrino emission from each spatiotemporal point of the model. Thus, we aim for the result
\begin{equation}
 I_{\nu}=I_{\nu}(\vec{\mathrm{r}},t)
\end{equation}
where the 3D space location $\vec{\mathrm{r}}$ is represented by the x, y, and z coordinates of the computational cell in question. Time t is obtained from the time tag of the hydrocode snapshot to which the cell belongs {(MHD datasets are four-dimensional, including the dimension of time)}. 
 The above equation is globally applied to all selected PLUTO data (the user may select beginning and end times for the global calculation). 

For reasons of economy, a double filter was applied, whereby neutrino emission was calculated only in cells in which the velocity is not too far from the LOS (cos(losu) greater than 0.08) and whose speed is at least 0.1 c. 

On the neutrino scale, the resulting jet's kinetic power was  L$_{\rm{k}}$ = 2 $\times$ 10$^{38}$ (see the ~{Appendix}
~\ref{energetics}). In \cite{Reynoso2009}, the authors argue a 10\% Eddington jet kinetic luminosity, leading to  L$_{\rm{k}}$ = 10$^{38}$ ergs$^{-1}$ for a 10 M$_{\odot}$ black hole, which is comparable to the current case. In \cite{Reynoso2009},  either $\frac{L_{\rm{p}}}{L_{\rm{e}}}\simeq 100$ or $\simeq$1 was adopted; we selected the former, which implies a hadronic jet. As an approximation, neutrino emission from a high-energy proton distribution is obtained, omitting  effects (such as synchrotron emission cooling) from the corresponding high-energy electron distribution in the jet. The jet base is situated near the center of a Cartesian grid. The same spatial resolution is employed on both the radio and  neutrino scales, resulting in a higher jet power in the radio-scale models. Nevertheless, this can be offset by a normalization process, while keeping densities, which strongly affect the emission and absorption calculations, realistic. 

The current work constitutes an improvement over previous work \cite{Smponias_2021}. More specifically,  time synthetic neutrino images are produced, leading to a more detailed view of the model system, as opposed to merely synthetic energy spectra in \cite{Smponias_2021}. Furthermore, the finite speeds of neutrino ``rays'' are now incorporated in the model, leading to more realistic neutrino imaging of  model jets. What is more, the current work focuses on sidereal emission from model jets, whereas \cite{Smponias_2021} provided a more general study of neutrino emission from jets viewed at different angles.

\subsection{Model Parameters}

 Table \ref{Table-for-run-data} shows a number of simulation parameters. These include the  computational cell length, jet density, and wind maximal densities (which  gradually decline away from their sources).  {\label{neutrino_density} 
 
On the neutrino scale, the jet kinetic luminosity is 2.5 $\times$ 10$^{38}$ erg/s, using a low spatial resolution of 60 $\times$ 100 $\times$ 50. Due to limited computing resources, such a low resolution was necessary in order to accommodate both the heavier neutrino emission calculation that followed and the time-resolved nature of the calculations, which essentially leads to four-dimensional datasets.} {\label{radio_density} 

In the radio scale model, the use of temporally resolved data, consuming computing resources, and the use of a fixed homogeneous grid, also necessitates a smaller spatial resolution, namely 60 $\times$ 100 $\times$ 50, and thus a larger computational cell. Radio-scale model jets are then more powerful, in order to keep their densities realistic, at the low resolution employed. At the jet nozzle, this means a thicker jet. Opting to keep jet density realistic, the overall kinetic luminosity then becomes higher than normal, mainly in the heavy jet radio model. Nevertheless, a normalisation process (electron power law constant K) may, to a certain extent, absorb the effects of an artificially increased jet power. On the other hand, synchrotron emission and absorption coefficients do depend, among other things, on the local density, so a need arises to focus on the density in this context}. 

In PLUTO, the piecewise linear method was set up using the MUSCL Hanckock integrator with an ideal equation of state. In the neutrino-scale simulation, the binary companion is located outside the grid, and was estimated to be at most up to an order of magnitude larger than the compact object. As mentioned above, the jet speed is~0.8c. 



\section{Results and Discussion}
\label{results}




\subsection{General}
The twin jet simulations in this paper represent three fiducial microquasars, dynamically set up to resemble a number of actual microquasars. In the synthetic imaging process (rlos code), a focused beam geometry was employed in combination with a finite ray speed. Synthetic images were projected  onto a fiducial imaging screen, parallel to the side of the grid (YZ), before the beam converged to a focal point, Figure \ref{focused_beam_imaging_geometry}.



In general, the imaging process may or may not use all snapshots available to it depending on the light crossing time of the model segment (adjusted through the {clight} 
 parameter in rlos, Appendix \ref{testingparameters}). Trying to read more snapshots than  are loaded corrupts the hydrocode time array of rlos, called T, resulting in errors.

For the neutrino calculation, as mentioned above, a double filter was used for the velocity and for the (los, u) angle. A minimal velocity and a  maximal angle were set in order to trigger the neutrino emission calculation for a particular cell. This way, the expensive part of the simulation was only performed when it was really worth it. This partly alleviated the discrepancy between computational costs of the dynamic and the radiative parts of the model. 

For the radio calculation, a similar filter was applied for the (los, B) angle. The use of filters highlights sideways emission from jet elements, with near-relativistic velocities roughly perpendicular to the jet axis.

An important aspect of this modeling approach is that a cell has a different visible emissivity  from Earth to that of its neighbors. This is because each cell may differ from the next one in terms of velocity value and orientation to us. Consequently, the output of a hydrodynamic model will differ from that of a steady-state model. Individual flow elements may appear either boosted or de-boosted depending on their velocity's orientation to the observer. 

Even from the side, adequate emission may be obtained due to elements of the flow moving relativistically sideways, especially during the initial   jet front expansion. Furthermore, delays in the arrival of emission emanating from the inner part of the jet mean that the initial violent interaction between the jet and the surrounding wind still affects synthetic images taken at subsequent time instants. Inner winds that are heavier than the jets, as was the case in the model runs, further prolong the effects of sidereal emission on the images by temporarily constraining the jet's ``bubble'' near the jet base. Thus, initial jet expansion still echoes in images taken later on in the simulation.

The distances in the synthetic images are in computational cells, where, in this work, one cell = two hydrocode length units. On the other hand, in the hydrocode data renderings, the scale is in hydrocode length units. For each scenario, the hydrocode length unit is shown in Table \ref{Table-for-run-data}.

\subsection{Radio-Scale Model Results}
\label{r-scale} 
 
\begin{figure}[H]
\includegraphics[width=11.5 cm]{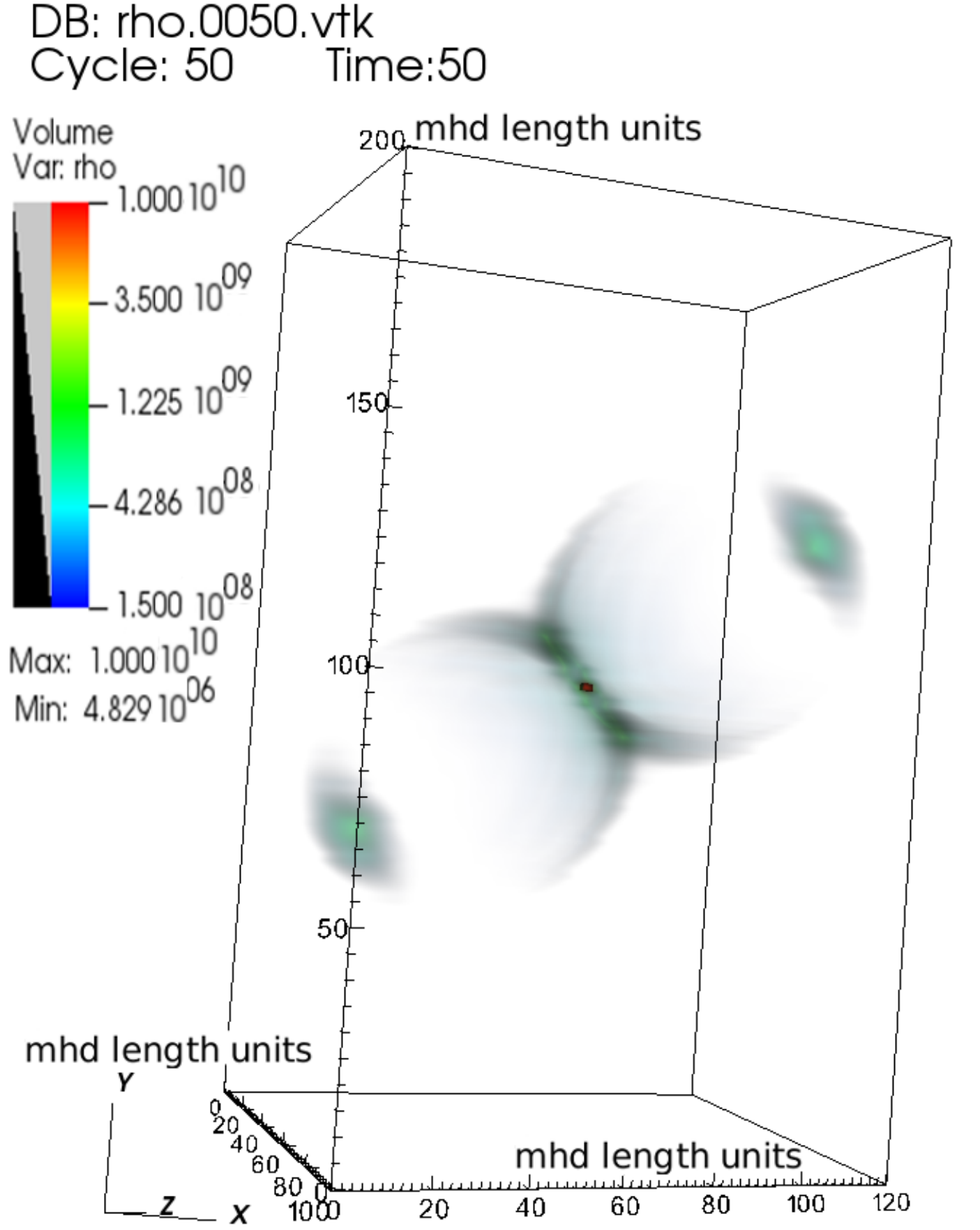}
\caption{{The radio-scale light jet model,} 
 shown at snapshot 50 (t = 100 ks). The plasmoids traverse the stellar wind, moving in opposite directions. The approaching jet is the one moving towards the beginning of the axes. The ambient wind is also proportionally lighter than in the heavy jet model, therefore resulting in similar dynamics to the heavier jet model. The plasmoids here also traverse a dense inner stellar wind, giving rise to an equatorial concentration of matter, detectable in the synthetic radio images. Overall, this lighter jet is more realistic in terms of densities and overall kinetic luminosity. In this figure, length units are taken from the MHD simulation, where two such units equal a computational cell in length. One MHD length unit here equals 10$^{13}$ cm (Table \ref{Table-for-run-data}). }
\label{radio_scale_jet_light}
\end{figure}

\begin{figure}[H]
\includegraphics[width=12.5 cm]{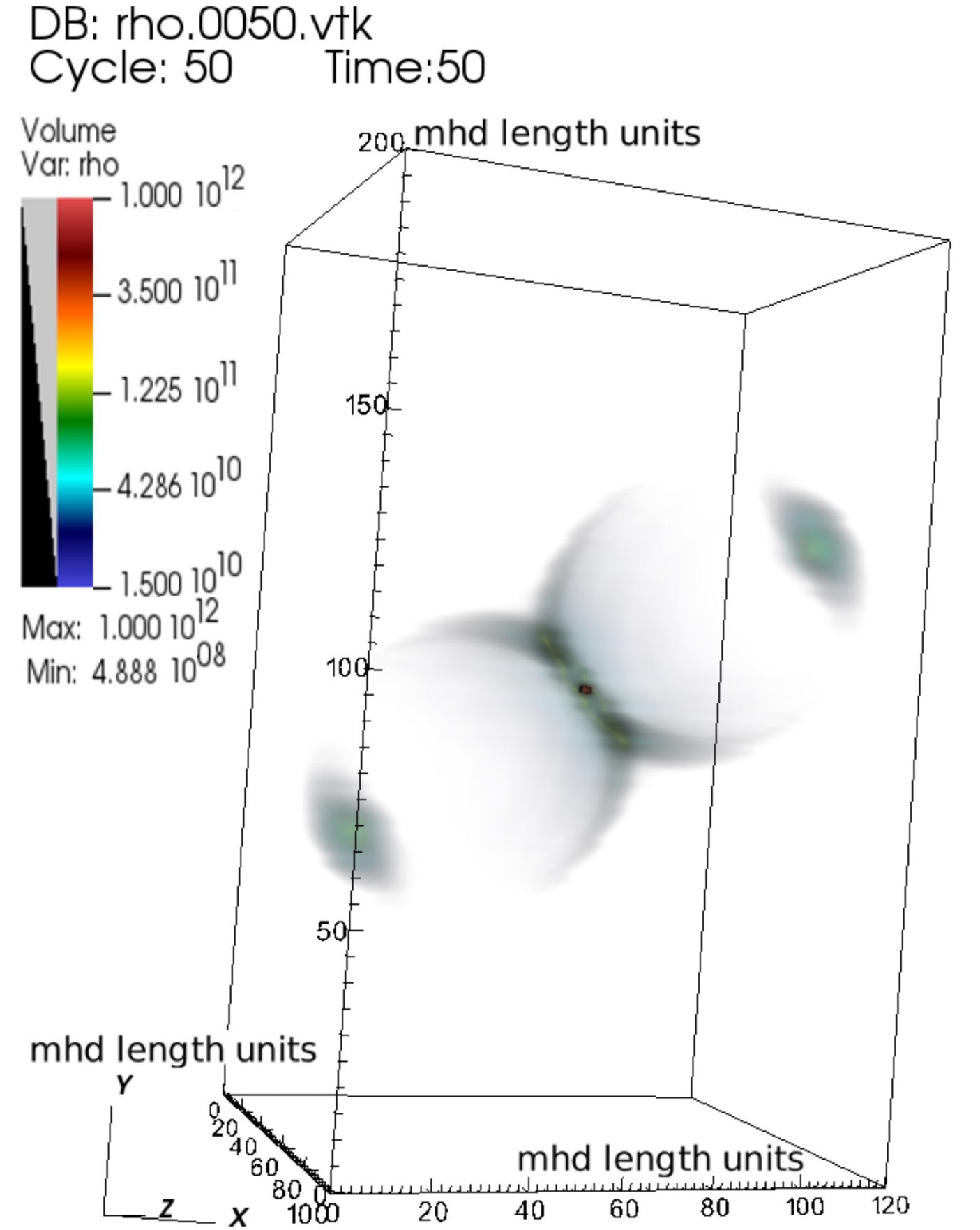}
\caption{{The radio-scale heavy jet model, shown at} 
 snapshot 50 (t = 100 ks). Blobs traverse the stellar wind, moving in opposite directions and forming a pair of intermittent jets. The approaching jet is the one moving towards the beginning of the axes. A cocoon formation appears, inflated by fast moving bolides, traversing the ambient wind of the companion star, which is denser than the jet near its base. Inner wind material is  pushed sideways by the jet, facilitating the creation of an equatorial zone, also detected in the synthetic radio images of the model system. In this figure,  MHD simulation length units are employed, where two such units equal a computational cell in length. An MHD length unit here equals 10$^{13}$ cm (Table \ref{Table-for-run-data}). }
\label{radio_scale_jet_heavy_visit}
\end{figure}

\textls[-25]{In the radio-scale model, ejected plasmoids form an intermittent twin jet  (\mbox{Figures \ref{radio_scale_jet_light} and \ref{radio_scale_jet_heavy_visit}})} and inflate a twin cocoon while traversing the companion {star}'s 
 stellar wind. 

Moving away from the jet base, plasmoids seem to expand while crossing ambient wind, thanks in part  to a declining wind density away from the binary system. A relatively low toroidal magnetic field further facilitates plasmoid expansion in the model. 

An equatorial concentration of wind, and possibly jet, matter forms dynamically (Figures \ref{radio_scale_jet_light} and \ref{radio_scale_jet_heavy_visit}), leading to some emission from that region, as seen in the corresponding synthetic radio images that follow.

\begin{figure}[H]
\includegraphics[width=9 cm]{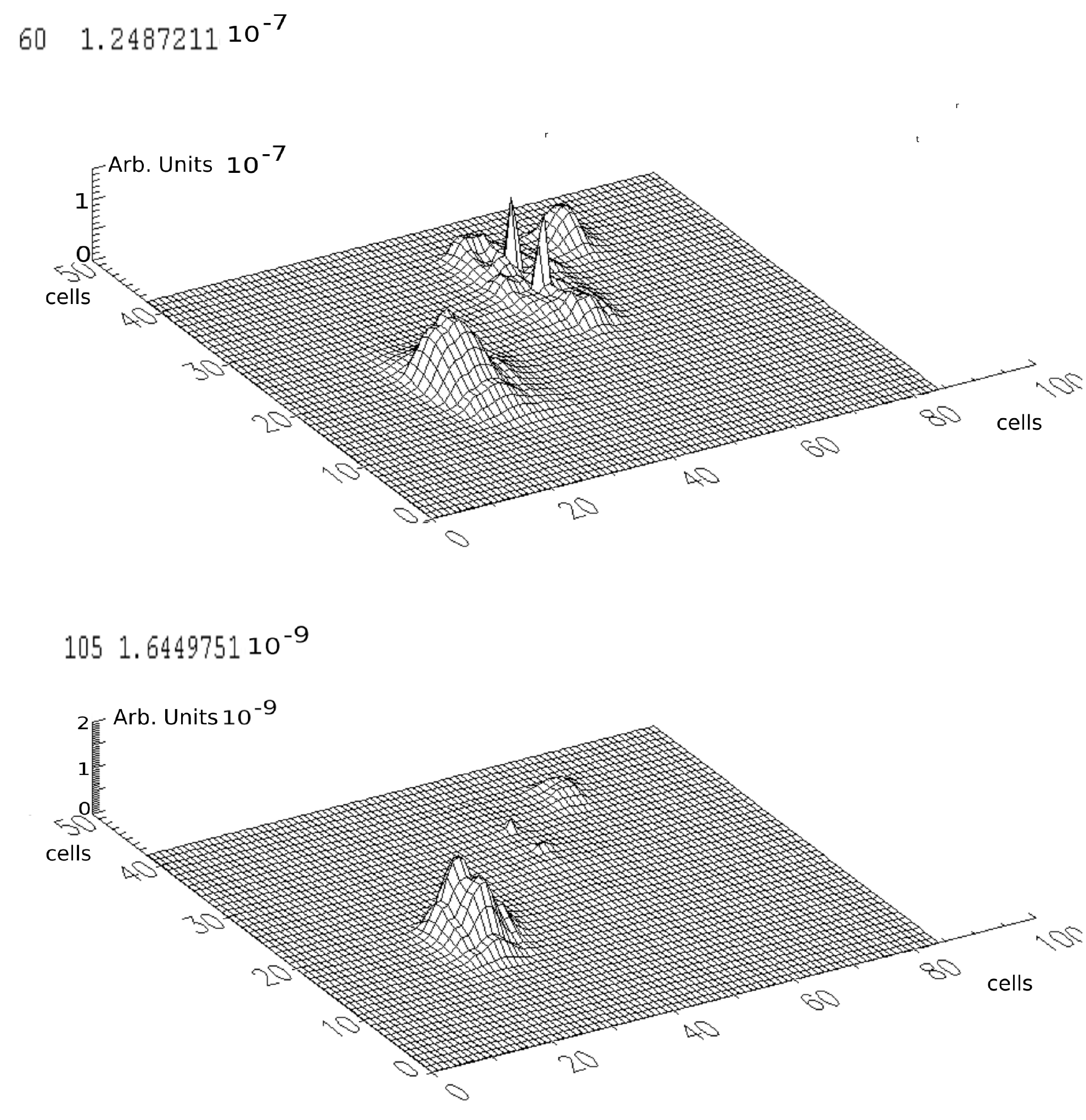}
\caption{{Unnormalized synthetic radio synchrotron} 
 images of the model system at 8 GHz. \textbf{Top}: the heavier jet model at snapshot 60 (t = 120 ks). \textbf{Bottom}: the lighter jet at snapshot 105 (t = 210 ks). In both cases, we can observe blobs moving in diametrically opposite directions. A finite c is taken into consideration in these plots, resulting in the approaching blob apparently moving much faster than the receding one. The approaching plasmoid is also much brighter, even though  the blobs in each pair are essentially the same in the hydrocode. Furthermore, the images shown here correspond to earlier blob locations in the hydrocode run due to the delay in the ray arrival to the fiducial observer. The total intensity in the heavy model is roughly two orders of magnitude larger than the lighter one, in rough proportion to the ratio of jet nozzle densities in the two cases. A stronger equatorial emission appears in the heavy jet case, attributed to the higher densities of this run, as well as to the earlier timetag of the heavy model synthetic image.}
\label{radio_synth_images}
\end{figure}

\begin{figure}[H]\vspace{-12pt}

\includegraphics[width=12.5 cm]{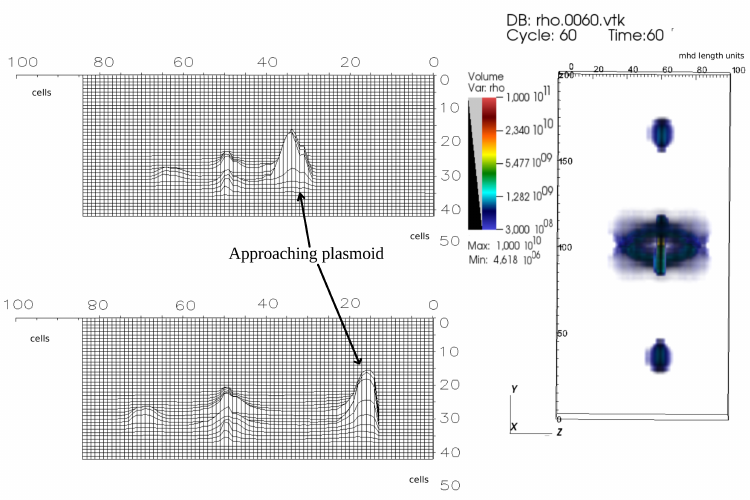}
\caption{\textls[-10]{Apparent superluminal motion study setup; the jet is at 55 degrees to the LOS with a speed of 0.8c. In the synthetic images (\textbf{top left-hand corner} and \textbf{bottom left-hand corner}), approaching and} }
\label{apparent_sl}
\end{figure}

\textls[-15]{We can observe the apparent acceleration of the approaching plasmoid (\mbox{Figures \ref{radio_synth_images} and \ref{apparent_sl}}}) on the sky plane (fiducial model screen), while the receding blob moves slower. The approaching blob is also brighter than the receding one. What is more, taking into consideration the finite nature of the speed of light,  earlier dynamics affect images taken at later times. Synthetic radio images of the model system in general exhibit a delay versus actual hydrocode plots bearing the same time tag {(Figures \ref{radio_synth_images} and \ref{apparent_sl}}). 

\begin{figure}[H]
\includegraphics[width=12.5 cm]{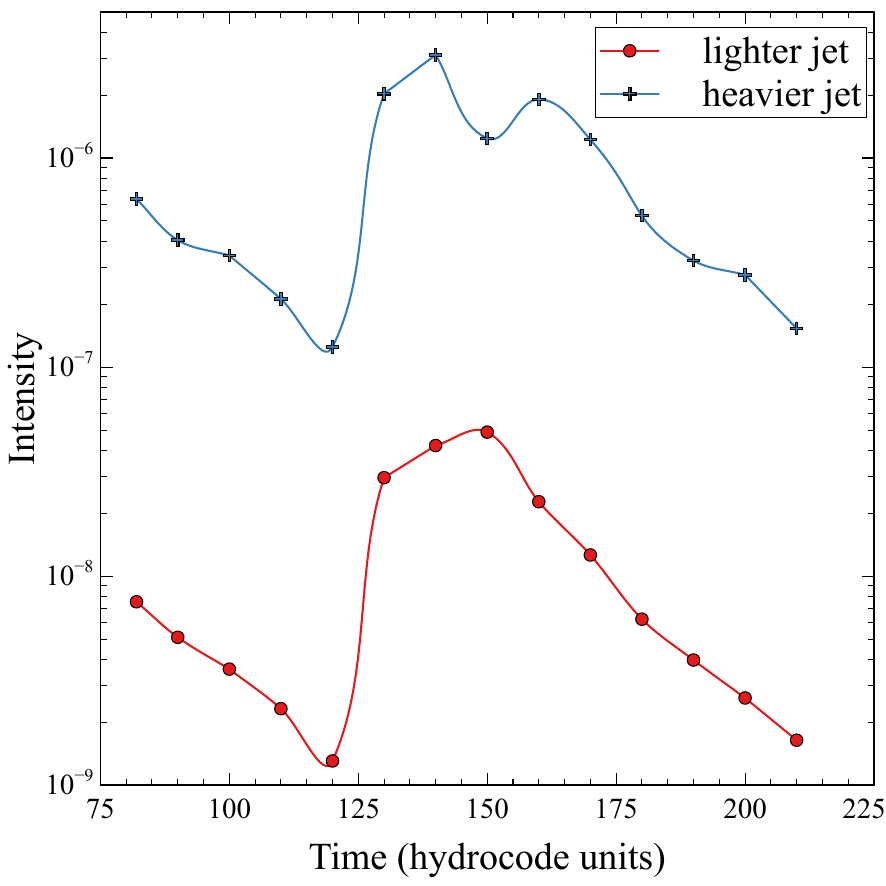}
\caption{Radio synchrotron model time curves, presenting the evolution of the total unnormalized intensity over the injection of a new plasmoid pair. Blob injection occurs during the first 30 out of every 100 time units. In this figure, the injection of the second pair of blobs is presented. The delay in their detection, of around 25 time units,  is attributed to the finite ray travel time from the model system to the fiducial imaging plane. In both cases, a faster rise is followed by a more gradual decline.  Nozzle densities: heavy jet, $\rho_{\rm{jet}}$ = 10$^{12}$ cm$^{-3}$; light jet, $\rho_{\rm{jet}}$ = 10$^{10}$ cm$^{-3}$.}
\label{both_heavy_and_light_radio_sync}
\end{figure}

A toroidal magnetic field threading the blobs leads to synchrotron emission in the direction of the observer, even from the receding blob (only two Doppler factors are employed to allow some visibility of receding plasmoids). The model   ejected plasmoid pair exhibits a quick rise in intensity and a more gradual decrease over time, Figure \ref{both_heavy_and_light_radio_sync}. In Figure  \ref{both_heavy_and_light_radio_sync}, for the heavy jet case (top curve), a previous pair of blobs happens to leave the computational grid at the time of new blob pair insertion, so a drop in intensity occurs there. This drop is observed at model time 150, but it actually took place shortly before the end of injection of the pair of blobs (since there is a recovery in intensity after the drop and thus injection continues after the drop), at around time 120--125 (injection of a second blob pair ends at model time 130). 

A suitable jet orientation setup has the potential for apparent superluminal motion of the approaching jet. This is visible in Figure \ref{radio_synth_images} and is particularly notable in Figure \ref{apparent_sl}.

\begin{figure}[H]
\includegraphics[width=12 cm]{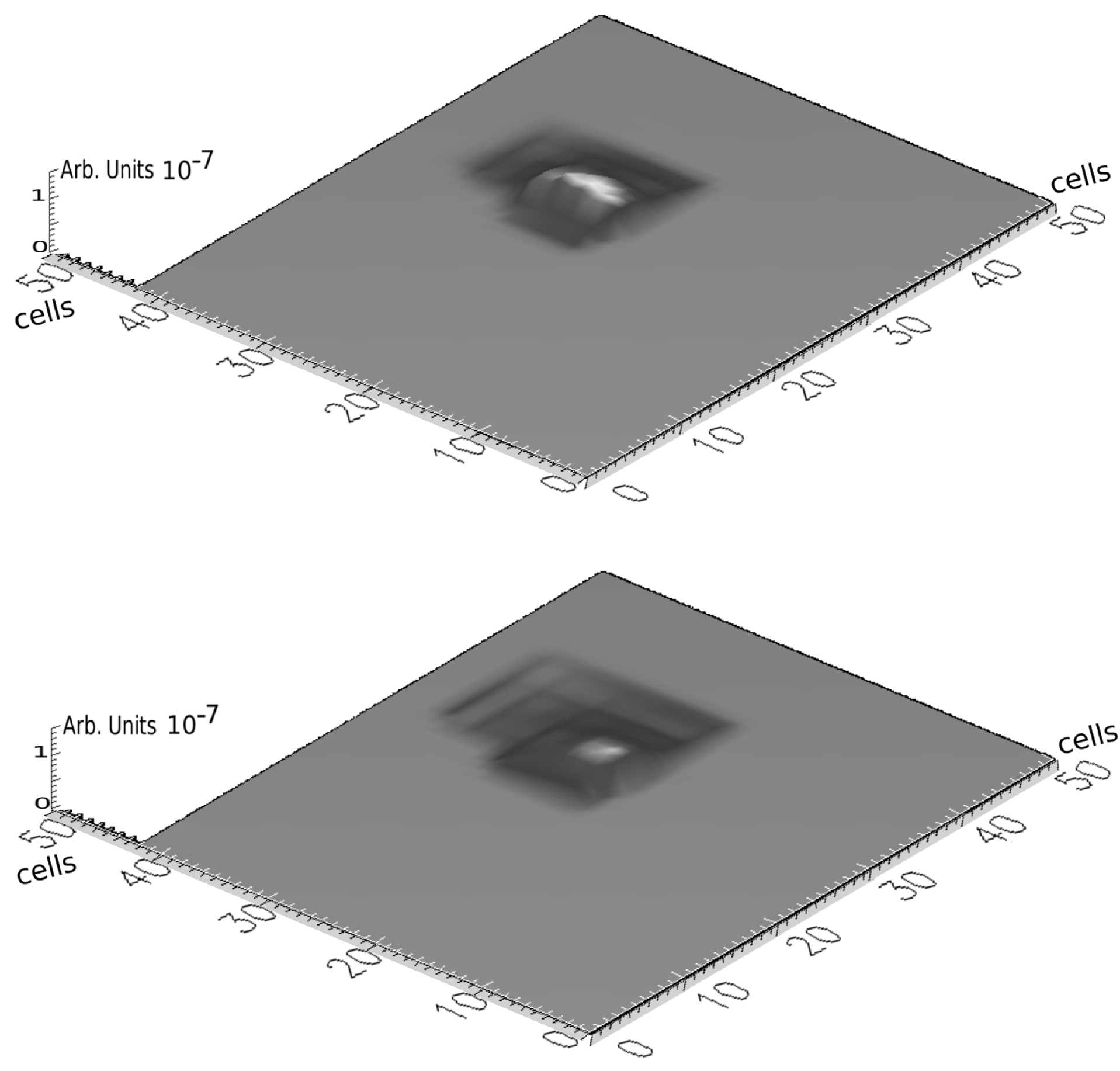}
\caption{{Synthetic model radio images of an approaching} 
 rectangular thin plasmoid. \textbf{Top}: the center of the rectangle has a higher intensity that its edges, attributed to a faster arrival of the central rays compared to  rays from the edges of the rectangle. The intensity peak is not fully symmetric within the inner rectangle as the imaging focal point is not exactly on the plasmoid's trajectory. On the other hand, by setting a speed of light c to  much higher than normal (\textbf{bottom}), this effect largely vanishes. We can also see shadows of the central region and of the receding twin jet blob. The vertical scale for intensity is  in linear, arbitrary units. }
\label{approaching_square_blob_rlos_and_visit}
\end{figure}

As a test, a simulation was also run with an approaching ``rectangular'' blob. A frontal synthetic image of the blob, moving straight into the fiducial observer (Figure \ref{approaching_square_blob_rlos_and_visit}), demonstrates the effects of a finite ray speed, as central rays from the rectangle appear stronger than those from its corners due to the delayed arrival of the latter.

\subsection{Neutrino-Scale Results}
\label{n-scale}
\begin{figure}[H]
\includegraphics[width=10.5 cm]{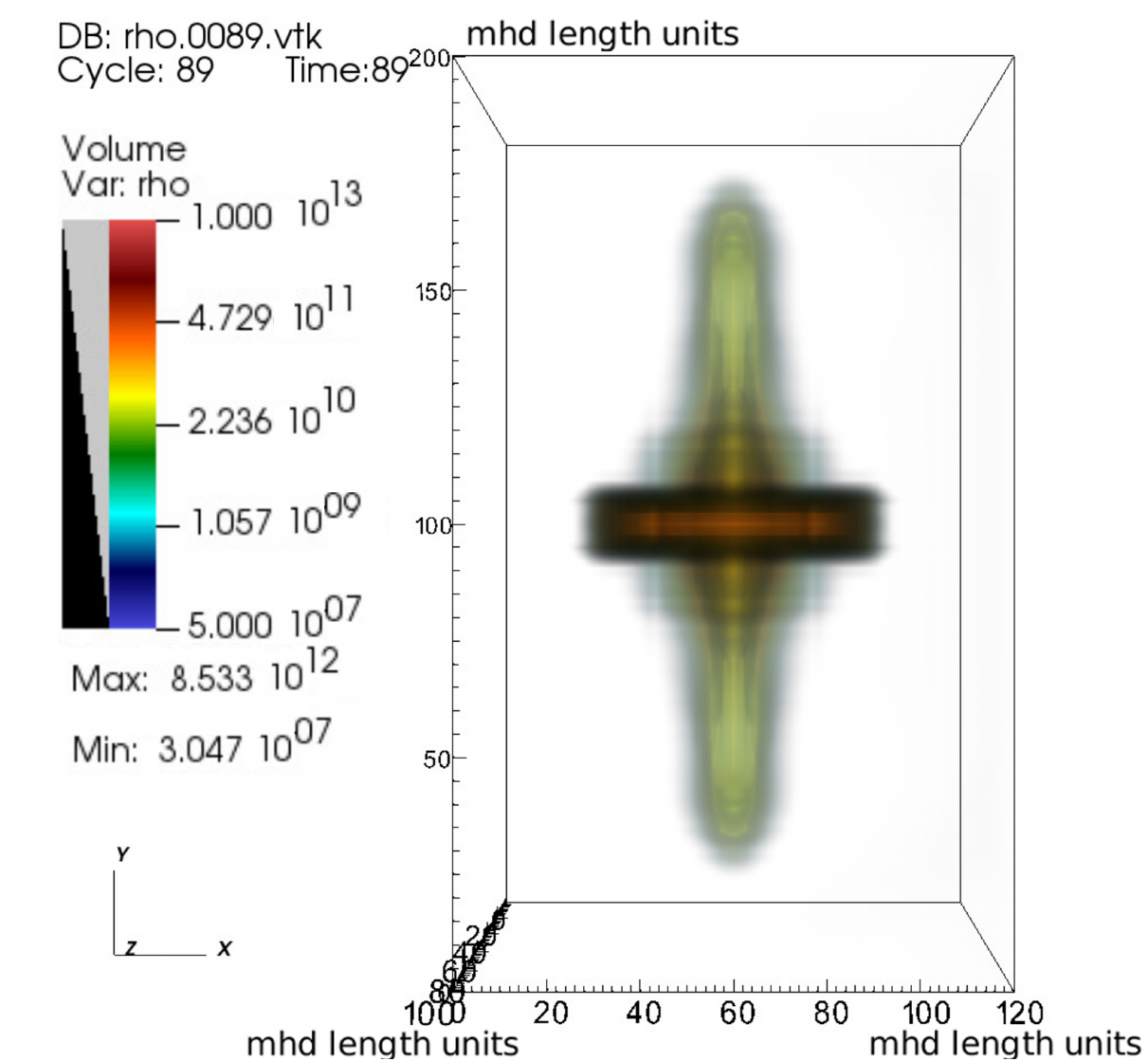}
\caption{{A continuous jet is shown} 
 on the neutrino emission scale, traversing the heavier accretion disk wind construct and also the companion star's stellar wind. Narrow collimated jets, lighter than the inner winds, propagate first through the accretion disk wind and then the stellar wind, thus inflating a ``composite'' cocoon structure. In this model, the accretion disk construct appears to divide the inner jet region into two separate dynamical environments, one for each jet.}
\label{neutrino_scale_jet}
\end{figure}

The PLUTO hydrocode was run in order to simulate the jets on the neutrino-emission scale (Figure \ref{neutrino_scale_jet}). A number of auxiliary PLUTO user parameters were also defined  beforehand in order to accommodate particle emission results from each cell later on. The purpose was to prompt PLUTO to create additional data files filled with a random number. These additional data files, while filled with meaningless data for now,  still form part of the PLUTO data save system. They are meant to act as a vehicle in order to accommodate particle emission results later on, with one  file for each neutrino energy.

The nemiss program was then run to calculate neutrino emissions from hydrocode data for a specific imaging geometry and setup. This program is able to read 4D spatiotemporal data outputs from PLUTO and convert them into a 5D array,  including the particle energy as the fifth dimension. Then, nemiss calculated the neutrino emission at each point of the 5D data array. Nemiss results were then saved into the aforementioned suitably prepared data files of the PLUTO hydrocode. Thus, nemiss processes the PLUTO output to add a neutrino emission spectrum at each spatiotemporal data point.

\begin{figure}[H]\vspace{-12pt}
\includegraphics[width=8.5 cm]{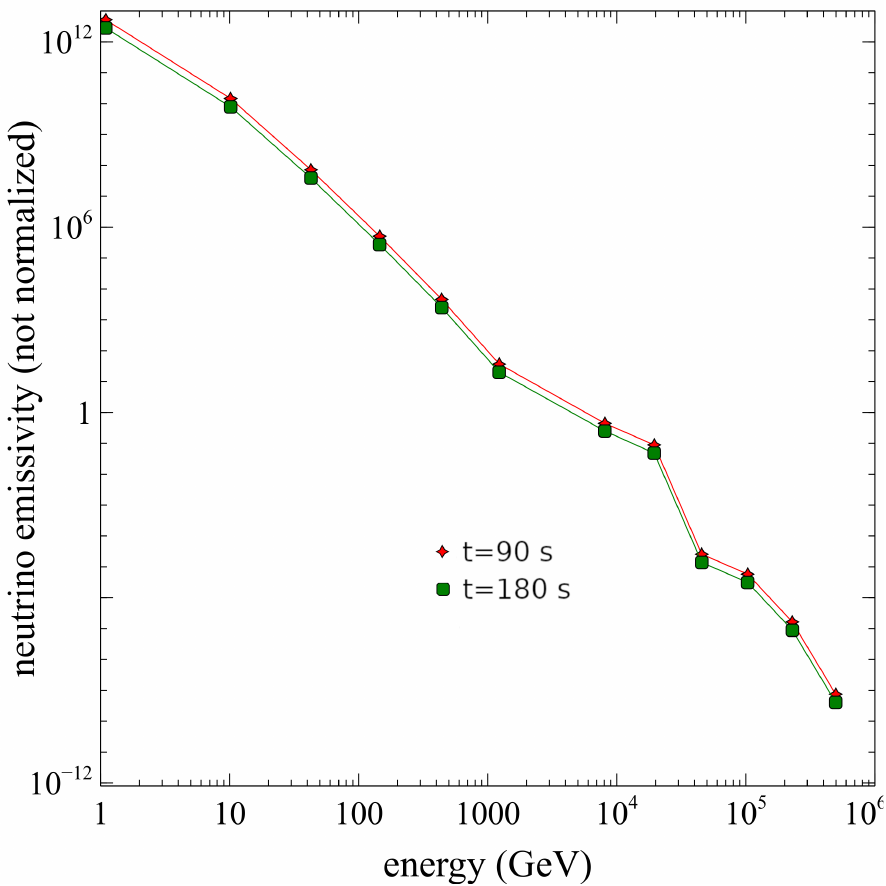}
\caption{{Neutrino unnormalized intensity } 
plots at snapshots 45 (t = 90 s, top) and 90 (t = 180 s, bottom). The sum of all  synthetic neutrino image pixels at each energy is plotted at each time instant. At the earlier time instant, the intensities are higher across the spectrum. This is attributed to intensity decreasing with time. The finite nature of the speed of light was taken into account when calculating these plots. Furthermore, at both time instants the intensity decreases with energy over an energy spectrum covering six orders of magnitude. In this model run, jet dynamics evolve smoothly, so the spectra remain very similar in~shape across the two instants here. }
\label{neutrino_both_plots}
\end{figure}

\begin{figure}[H]
\includegraphics[width=10.5 cm]{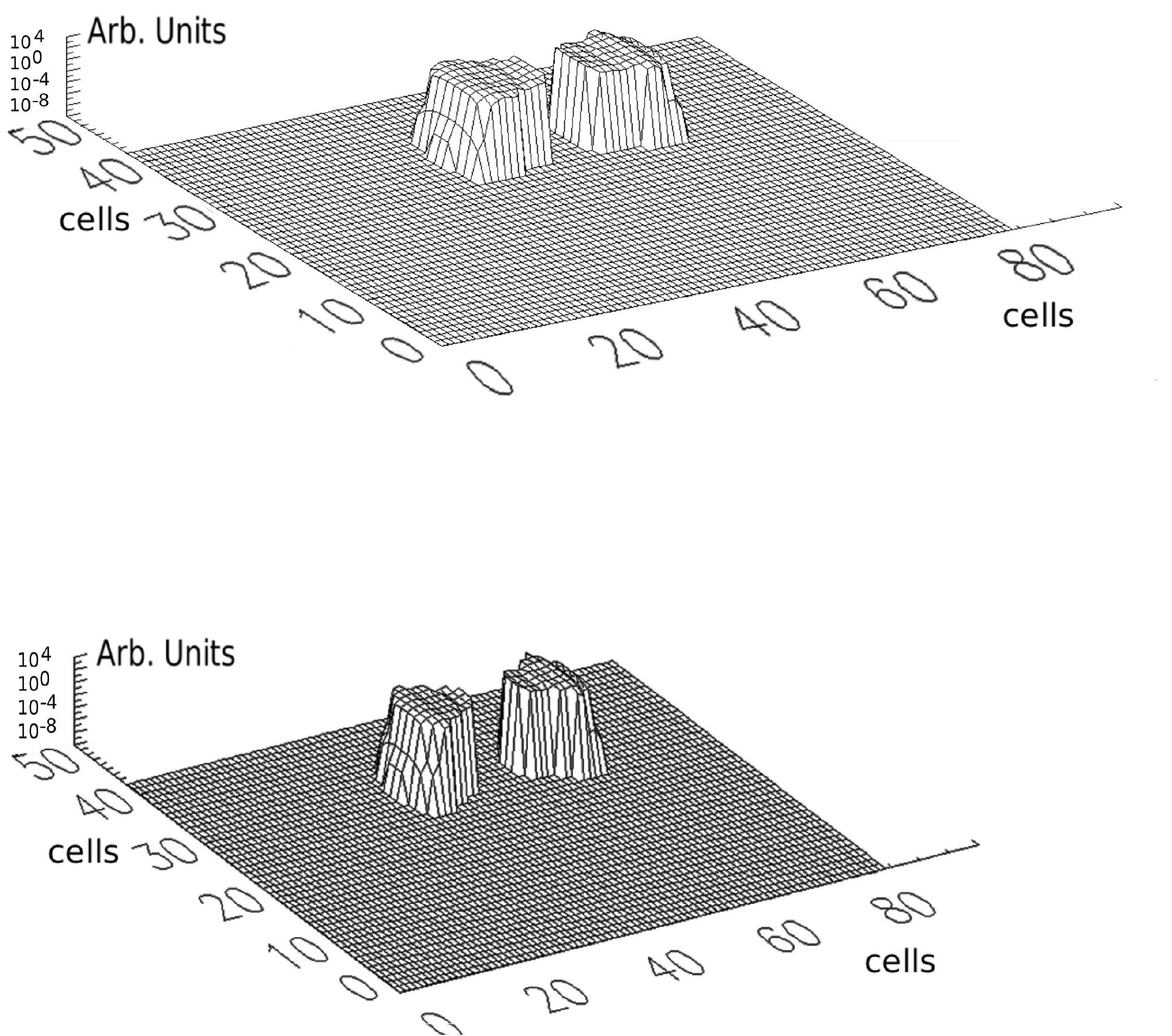}
\caption{{Unnormalized neutrino intensity images} 
  of the same hydrocode run at snapshots 45 (\textbf{top}; t = 90 s) and 90 (\textbf{bottom}; \mbox{t = 180 s}) at a specific energy (400 GeV). In order to produce the image in rlos, the focused beam imaging method, with back in time integration along the LOS, was employed. In all other energy slots of the spectral energy distribution, the image shape is quite similar, though not same, but the intensity scale does fall with energy. Emission calculations were double-filtered by setting a maximum   angle between the local LOS and local u and a minimum  local velocity.}
\label{two_unnormalized_synthetic_neutrino_images_at_different_times}
\end{figure}

PLUTO data processed by nemiss were then ready to be read by the relativistic imaging code rlos \cite{RLOS}, which produced synthetic neutrino images of the system. The sum of intensities  over all  synthetic jet images was calculated over a range of particle energies, leading to a plot of jet neutrino intensities, Figure \ref{neutrino_both_plots}. Furthermore, the model was run at two different time instants: shot number 45 (t = 90 s) and shot number 90 (t = 180 s). A relativistic imaging process was used to produce synthetic images, Figure \ref{two_unnormalized_synthetic_neutrino_images_at_different_times}.

\begin{figure}[H]
\includegraphics[width=10.5 cm]{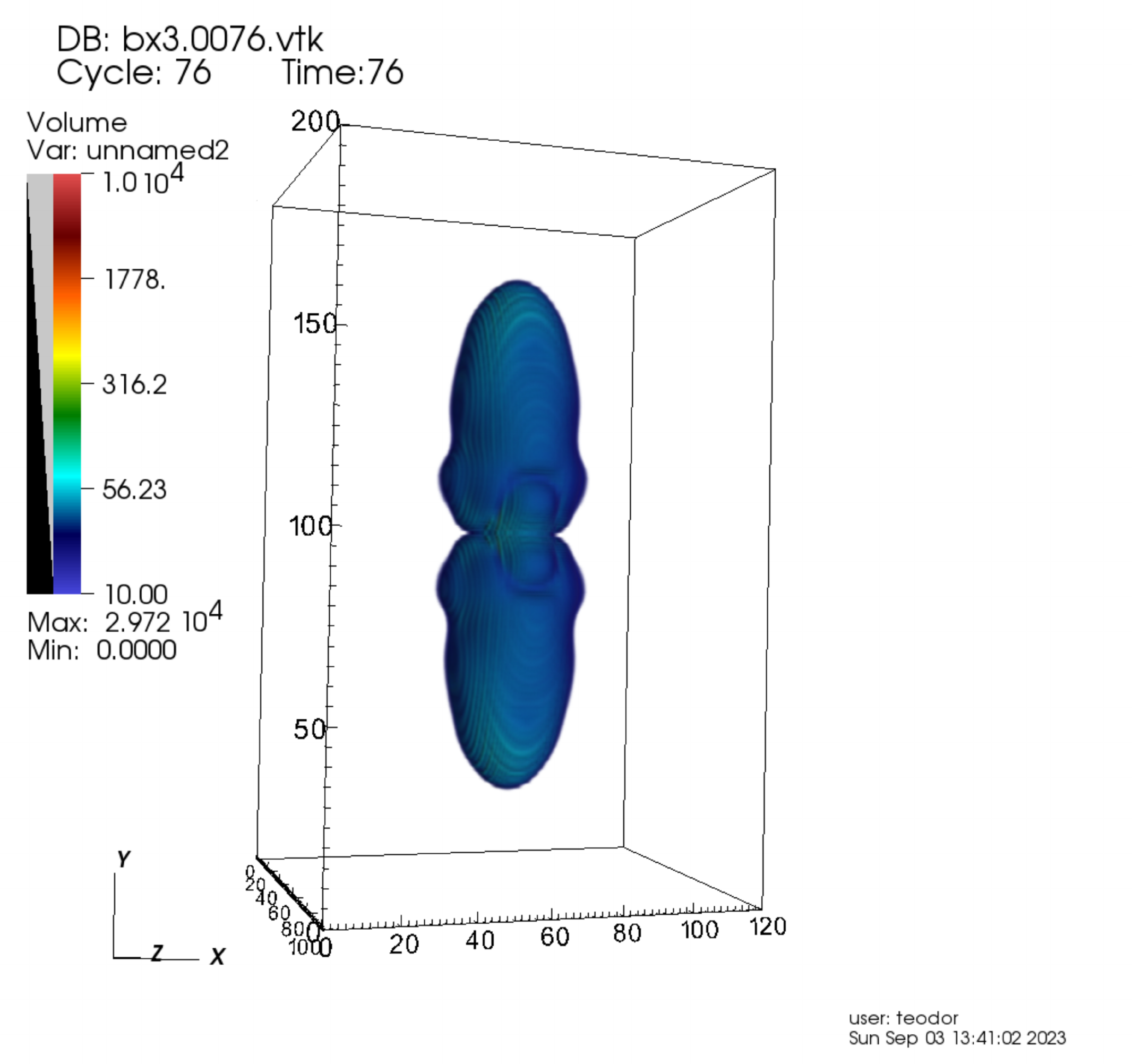}
\caption{{The absolute value of the } 
toroidal magnetic field component of the continuous jet is shown on the neutrino emission scale. The scale is logarithmic and the field appears to thread the jet, contributing to its confinement along its path. The field also plays an important role in the neutrino emission calculations.}
\label{neutrino_scale_bfield}
\end{figure}

Figure \ref{neutrino_scale_jet} shows narrow jets that are slowly expanding into the surrounding winds, collimated by a strong toroidal magnetic field component (Figure \ref{neutrino_scale_bfield}). This small half-angle is then rather counter-intuitively expected to result in a faster decline in neutrino emission with energy, as discussed in  \cite{Reynoso2009}. Jets interact with winds, their cocoon expanding sideways as well as forwards into the surrounding wind matter.  Furthermore, sidereal expansion of the inner part of the cocoon accrues enough dense and relatively fast matter to achieve adequate (beamed, according to \cite{TR11}) sideways neutrino emission, as seen in the synthetic neutrino images that follow. The effects of the accretion disk and its wind construct therefore seem to play an important dynamical role in providing sideways particle emission from the jet system. This is important for sideways neutrino emission (Figure \ref{two_unnormalized_synthetic_neutrino_images_at_different_times}) because it shows localized neutrino emission along a direction perpendicular to the jet axis.



%
%
%
%



As mentioned earlier, adequate neutrino emission may occur towards Earth, even from MQ jets not aligned with the LOS to Earth. This leads to a rather increased number of MQ candidates for neutrino detection, particularly those with rich dynamic interactions with the surrounding winds.

The above result positively affects emission from a galactic MQ distribution \cite{Smponias_2021}. Even in MQs whose jets point perpendicularly to the LOS to Earth, some relativistic boosting may still appear in parts of the flow moving towards us, especially early on in the ejection~event.



\begin{figure}[H]
\includegraphics[width=10.5 cm]{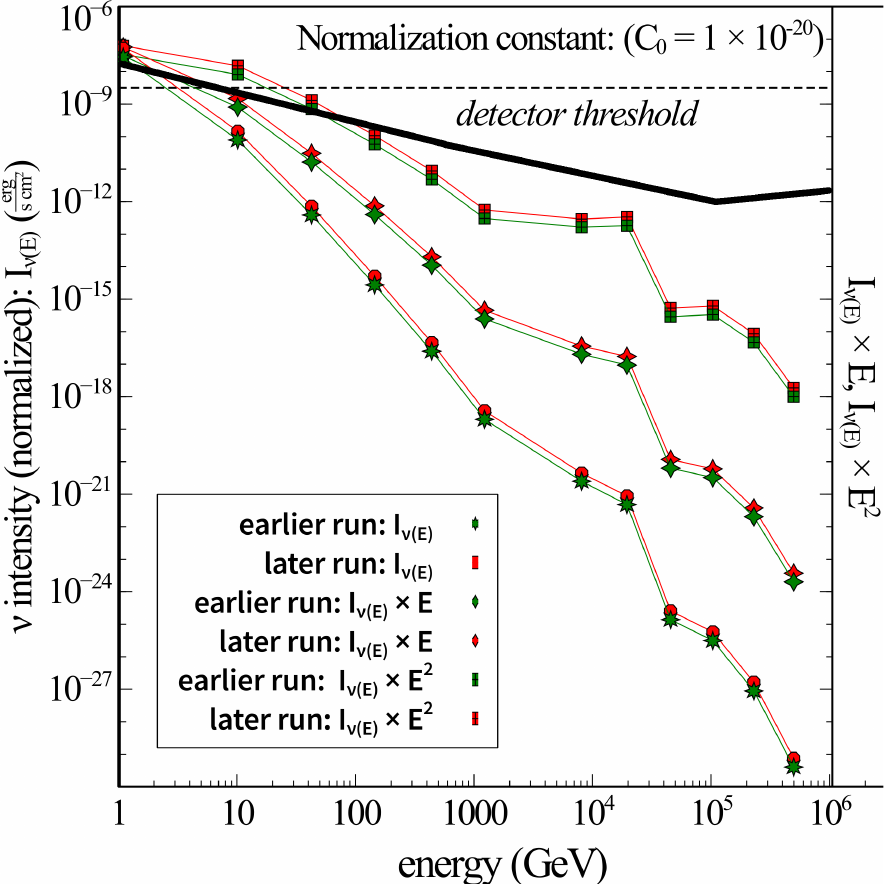}
\caption{{Neutrino-normalized intensity} 
 spectral emission distribution plots at snapshots 45 and 90 (t = 90 s and 180 s, respectively). The synthetic imaging viewing angle of the model system used to produce these spectra is from the side, roughly perpendicular to the jets. The contribution to the emission is  mostly from matter elements moving sideways during jet cocoon expansion through ambient winds. Sizeable sideways emission does appear in the synthetic images, and that is reflected in the normalization process assumptions used here. It should be noted that the quantitative intensity difference between the earlier and later snapshots is due to the actual dynamical evolution of the jet system in the PLUTO RMHD simulation, as both spectra were produced using the exact same normalisation assumptions. The two spectra exhibit similarities, as expected by the similarity of the synthetic images in Figure \ref{two_unnormalized_synthetic_neutrino_images_at_different_times}. We can also see the reference sensitivity of a cubic km detector array in the first approximation (dotted line), and an improved version depending on the energy (thick line)~\cite{Aharonian_2018}.} 
\label{neutrino_both_plots_norm}
\end{figure}

As shown in \cite{Smponias_2021}, the scale of total emission is expected to increase the closer the LOS is to the approaching jet axis. The current results (Figure \ref{neutrino_both_plots_norm}) demonstrate the possibility of potential observations, even when the jet is observed from the side. In other words, in this work, beaming \cite{TR11} is localized in cells whose relativistic speed points towards Earth, even though the jet axis lies essentially perpendicular to our line-of-sight. As a concrete example, the detection ability of a cubic km array is depicted in a normalized SED plot, Figure \ref{neutrino_both_plots_norm}, presenting a reasonable possibility for detection.




%

The above estimate may then be employed  to provide a rough estimate of expected neutrino emission from a microquasar distribution   in the galaxy. The authors of \cite{Paredes2003} proposed an estimated population of around a hundred systems in our galaxy. Furthermore, their discussion of $\gamma$ ray emission from microquasars clarified the importance of relativistic boosting in jet emission. Thus, orientation to Earth plays a major role here, and the situation is similar for neutrino emission \cite{TR11}.

{\captionof*{figure}{receding blobs are shown for a ray speed of $\simeq$200c (\textbf{top left}) at shot number 30, and for normal c (\textbf{bottom left}) at shot number 60 (the synthetic image view is rotated around the z axis by 180 degrees for visual clarity). The difference in timing between the two synthetic images represents an attempt to match the time delay of the normal c image to the single-shot image  at 200c. Distances in the synthetic images are in computational cells, where one cell equals two hydrocode length units. Top left: no real difference exists between approaching and receding plasmoids, resembling the corresponding hydrocode density plot to the right. Bottom left: the approaching blob  on the fiducial screen (sky plane) appears to be much faster than the receding blob; a clear demonstration of the apparent acceleration effect. On their right-hand side, a hydrocode data rendering of the same model run is shown; the scale is in hydrocode length units. A hydrocode density plot is also shown, demonstrating the inherently symmetric motion of approaching and receding hydrocode blobson the fiducial sky plane  (without using an imaging code).}}

\newpage

For neutrino emission, in a manner similar to \cite{Smponias_2021}, we proceed by assuming 100~systems at various distances ranging from a minimum of 1 kpc to a maximum of 30 kpc, with  an average kinetic luminosity similar to our model system. The linear dependence of emission on the latter quantity facilitates such a simplification. A distance of 1 kpc commands a flux at Earth of 25 times more than our model value, whereas the representative system situated at 30 kpc has 36 times less than at 5 kpc.

In this analysis, the angle of the line-of-sight to the x axis is employed to ensure compatibility with the rlos geometry setup. This angle is complementary to the angle to the jet axis.

 An orientation of less than 60 degrees might lead to an intensity an order of magnitude lower than our value, but a jet system aimed towards us could perhaps have up to 100~times more visibility on Earth \cite{Smponias_2021}. The latter point can be somehow altered by the present analysis insofar as, even from the side, the expanding cocoons lead to local sidereal beaming towards~Earth. 
 
\vspace{-6pt}

\vspace{-12pt}

\vspace{-6pt}

\vspace{-6pt}

Orientation then remains the most important factor, but the low intensity end of the range might be updated upwards, since the local cell velocity might still point towards Earth even if the jet axis does not. Then, we have the distance, and lastly, the jet kinetic~power. 

The above order of importance allows for an estimate of perhaps 5\% or, in other words, five systems with a very high relativistic boosting towards us. Then, we have  maybe 40 or 50 sources aimed at angles above 45 degrees but below 90 degrees, and lastly maybe 50~sources at below 45 degrees which still may have quite sizeable contributions. The high boost sources probably contribute the most on average, and the ones viewed from the side have a smaller effect. A possible system located at a smaller distance, oriented towards Earth, would of course dominate the above distribution, but the possibility of such an occurrence is not very large.

On the basis of the above discussion, we then use a rough average for a neutrino-emitting galactic microquasar located at 15 kpc, with the kinetic luminosity of our model (less affecting factor) and orientated at 37.5 degrees from the line-of-sight.

The reason for using an average angle of 37.5 degrees, higher than the estimate of 30 degrees in \cite{Smponias_2021}, is the  higher contribution from systems not aimed towards us due to increased sidereal emission from dynamical jet systems.  



The orientation of  the local velocity and the  magnetic field seems to play a crucial role in both neutrino and radio emission calculations. Consequently, the differential projection in each cell strongly affects the final images. 



It seems possible, even more than in \cite{Smponias_2021}, that the detection of background emission from a potential distribution of microquasars in the galaxy lies within the realm of modern detector arrays. This is also a consideration for the next generation of new or upgraded arrays. On the other hand, a single X-ray binary system might act as a galactic source of high-energy neutrinos. This is a potential target for a particle sensor with an increased angular accuracy. The variability of microquasars within the human timescale, combined with their relative stability as a known point source, offers a good target for observation, especially combined with sensors operating in the electromagnetic spectrum.

\section{Conclusions}
\label{conclusions}

In the radio-scale model, apparent acceleration of the approaching plasmoid is observed on the fiducial sky plane, as well as an increased brightness. On the other hand, the receding plasmoid appears to move slower, while also appearing dimmer. In the RMHD hydrocode model, each   blob pair is essentially identical, so the imaging model can capture the apparent relativistic acceleration and beaming at each jet point. Furthermore, the frequency shift is also included separately for each computational cell. This way, a more realistic simulation of the model  system  is achieved, facilitating a comparison with observations. The dynamics in the hydrocode and the magnetic fields threading the jets are then better connected to  the actual appearance of the system  to a detector array. 

In the neutrino-scale model, an updated result is obtained, favoring sideways emission from relativistic jets. This may lead to an increased number of MQ candidate neutrino~sources.

The  synthetic imaging method using special relativistic methodology may be expanded to a general relativistic framework. At every grid point, a gravitational potential may alter the course of a ray, depending on the local matter and energy properties, as provided by a suitable hydrocode simulation. This way, a broader range of astrophysical problems can be approached with increased realism. Particle emissions can be included, if suitable transformation equations are provided for their energy spectra from the source reference system to the stationary system.

\funding{{This research received no external funding.}} 

\institutionalreview{Not applicable.} 




\acknowledgments{{We are grateful to our colleagues} 
 for their valuable comments on the manuscript. Special thanks go to  G. E. Romero (UNLP, IAR) for their suggestions on improving the content of this~work.}

\conflictsofinterest{The author declare no conflict of interest.} 

\appendixtitles{yes} 
\appendixstart
\appendix
\section{Jet Size and Jet Energetics}
\label{energetics}

\subsection{Model Space Size}
\label{sizing}

The computational grid size is l$_{\mathrm{grid}}$ = 200 $\times$ 10$^{13}$ cm = 2 $\times$ 10$^{15}$ cm. Distance to the generic MQ model system is taken as D = 5 kpc, or D $\simeq$ 1.5 $\times$ 10$^{22}$ cm. Therefore, \mbox{sin($\frac{ \mathrm{l}_{\mathrm{grid}} }{ \mathrm{D} }$) $\simeq$ 1.3 $\times$ 10$^{-7}$}. For such small angles, sin(angle) $\simeq$ angle. Thus, arc size is around 1.3 $\times$ 10$^{-7}$ rad, or roughly 27 mas. The latter size is for the full computational domain, including jet and counterjet, as well as certain empty space margin. Consequently, for the core of the model grid, the model jet and counter-jet region spans no more than perhaps 15 mas. We may then consider a synchrotron emission region of up to 10 mas for a model jet, which is generally thought to be the inner compact region, where a flat to inverted spectrum may~occur.

\subsection{Normalization}
\label{normalization}

The jet kinetic energy at its base is \cite{Reynoso2009} 

\begin{equation}
E_{\rm{k}}= (\Gamma-1) m c^{2} 
\end{equation}
{A constant} 
 jet speed u is considered for this calculation, meaning a constant $\Gamma$ Lorentz factor. m is the mass of a jet portion traversing a jet cross section near the base. Then, jet kinetic power P$_{\rm{k}}$ is the kinetic energy traversing the cross section per unit time
\begin{equation}
P_{\rm{k}}=dE_{\rm{k}}/dt = (\Gamma-1) (dm/dt) c^{2} 
\end{equation}
where the speed is taken to be constant during an ejection episode (it was also set to be constant in the simulation described here). However, 
\begin{equation}
dm/dt=\rho dV/dt = \rho A dx/dt =\rho A u
\end{equation}
where $\rho$ is jet density, V is the volume of a thin jet cross section, of length dx, near the jet base, and A is the jet base cross section area, also taken  as a constant both in the simulation and here. Therefore,
\begin{equation}
P_{\rm{k}}=dE_{\rm{k}}/dt= ((\Gamma-1) \rho A) c^{2}u 
\end{equation}
or
\begin{equation}
P_{\rm{k}}=dE_{\rm{k}}/dt=  ((\Gamma-1) \rho N_{\mathrm{cell}} L^{2}_{\mathrm{cell}}) c^{2}u 
\end{equation}
where $A=N_{\mathrm{cell}} L^{2}_{\mathrm{cell}}$ is the area of jet cross section near the base, $L_{\rm{cell}}$ is the length of the edges of a cubical computational cell, at the jet base, $L^{2}_{\rm{cell}}$ is the frontal area of a cell, and $N_{\mathrm{cell}}$ is the number of such cells forming the jet base cross section.

We then express density as a function of proton number density $N_{\rm{p}}$ and proton mass~$m_{\rm{p}}$
\begin{equation}
\rho=N_{\rm{p}} m_{\rm{p}}.
\end{equation}

The field of view (FOV) of the synthetic image is taken at an estimated 1/10 of the total solid angle. In-there, a similar portion of total emission is assumed, allowing for the majority of emission to go in the vicinity of the jet axis orientation. In total, a field-of-view factor of 1/100 of total emmission is employed: f$_{\rm{fov}}$ = 0.01.

For the neutrino-scale model, let us define neutrino luminosity L$_{\nu}$ as the power emitted through neutrinos from the jet, which is a fraction $\alpha$ of the total kinetic jet power (jet kinetic luminosity P$_{\rm{k}}$ = L$_{\rm{k}}$).

 Consequently, $\alpha=L_{\nu}/L_{\rm{k}}$, represents the portion of jet power (total power of hot and cold protons) emitted in neutrinos (the latter portion, $\alpha$, is different than the portion of the hot proton energy converted to neutrinos, which usually bears a value of the order of a few percent). For normalisation, a working value is taken as $\alpha$ = 10$^{-3}$. This is compatible with a value of q$_{\rm{rel}}$ = 0.1, for the energy portion, q, of hot protons in the jet \cite{Reynoso2009,Reynoso2019}. It is out of energy portion q that a few percent end up to neutrinos. In summary, as an order of magnitude approximation, if jet kinetic energy is 100, then hot proton energy is taken as 10, and neutrino energy is taken as 1 percent of that 10, which is 0.1.  

The shape of the spectrum is also affected by acceleration efficiency \cite{Reynoso2019}, and from the opening angle of the jet \cite{Reynoso2009}, thus affecting the area under the neutrino spectrum plot. As an approximation for the above effects, we adopted a value of 0.01 for the energy transfer from nonthermal protons to the neutrinos. 

We also set u = $\beta$c. A less-than-unity positive filtering factor $f_{\rm{f}}$ is employed that  accounts for not using all jet cells, but only those with velocity orientation closer to the LOS and  with speed above a given limit. 
We then have
\begin{equation}
L_{\nu}=\alpha L_{\rm{k}}=\alpha P_{\rm{k}}=\alpha dE_{\rm{k}}/dt=f_{\rm{fov}} f_{\rm{f}}  \alpha (\Gamma-1) (N_{\rm{p}} m_{\rm{p}} N_{\mathrm{cell}} L^{2}_{\mathrm{cell}}) \beta c^{3} 
\end{equation}
{The intensity} of the jet is then expressed as $I_{\nu}=L_{\nu}/4 \pi D^{2}$, where D is the distance to Earth. Thus,
\begin{equation}
I_{\nu}=f_{\rm{fov}} f_{\rm{f}} \frac{1}{4 \pi D^{2}} \alpha  (\Gamma-1) (N_{\rm{p}} m_{\rm{p}} N_{\mathrm{cell}} L^{2} _{\mathrm{cell}}) \beta c^{3} 
\end{equation}

 
In the neutrino-scale simulation, the jet beam travels at $\beta$ = $\frac{u}{c}$ = 0.8, with a density of 10$^{11}$ protons/cm$^{3}$. L$_{\mathrm{cell}}$ is 10$^{10}$ cm, while the number of cells comprising the beam at its base at this resolution is N$_{\mathrm{cell}} \simeq$ 15. Distance to Earth is taken here with a typical value of D = 5 kpc or approximately 2 $\times$ 10$^{22}$ cm. 

We then integrate the area under the curve of the arbitrary units plot (Figure \ref{neutrino_both_plots}) for our case of viewing the jet from the side. That case is supposed, for the purposes of normalization, to be the one matching the orientation of the hypothetical system in relation to Earth. We perform a cumulative sum over the roughly 10 points. Thus, we find about 10$^{11}$, in arbitrary units (AU)*GeV. We replace an AU with a constant C$_{\rm{0}}$, so that \mbox{AU = C$_{\rm{0}}$ erg/(s*cm$^{2}$)}. 
We set $I_{\nu}$ = $L_{\nu}/4 \pi D^{2}$ equal to the area under the plot of Figure~\ref{neutrino_both_plots}, called PLOTAREA, expressed in units of C$_{0}$, in order to find the latter (normalization~constant)\vspace{-12pt}
\begin{adjustwidth}{-\extralength}{0cm}
\centering 
\begin{equation}
I_{\nu}= f_{\rm{fov}}f_{\rm{f}} \frac{1}{4 \pi D^{2}} \alpha  ((\Gamma-1) (N_{\rm{p}} m_{\rm{p}}) N_{\mathrm{cell}} L^{2} _{\mathrm{cell}}) \beta c^{3}= (\mathrm{PLOT AREA})*C_{0} \, \mathrm{erg/(s \, cm^{2}) \, GeV}
\end{equation}
\end{adjustwidth}
{For} our case, we find C$_{\rm{0}}$ $\simeq$ 10 $^{-20}$, which is the value of the arbitrary unit C$_{\rm{0}}$. Using the above constant, we multiply by it the value given in arbitrary units for the particle emission. Thus, the intensity plot is multiplied, and we arrive to the updated plot in Figure \ref{neutrino_both_plots_norm}, which may be directly compared to other models and to observations.

For the radio-scale model, we may consider an estimate of the intensity in radio at 8~GHz. Relativistic beaming is now present, for the E/M radiation. In a manner similar to particles (see above), beaming leads to an approximate value for the field of view factor of f$_{\rm{fov}}$ = 0.01. The portion of the jet kinetic power L$_{\rm{k}}$, emitted at the radio band in 8 GHz is estimated as PORTION. 

For L$_{\rm{k}}$ = 10$^{38}$ ergs$^{-1}$, L$_{\rm{k}} \times$ PORTION = (Power)$_{\mathrm{8Ghz}}$ = 4$\pi$D$^{2}$ $\times$ I$_{(\mathrm{Earth(8GHz)})}$. Adopting 1~Jy (1 Jy = 10$^{-23}$ ergs$^{-1}$cm$^{-2}$Hz$^{-1}$) as broadband intensity I at Earth, and a channel width, at 8 GHz, of 10$^{-2}$ GHz (10$^{-2}$ GHz = 10 MHz = 10$^{7}$ Hz), I$_{(\mathrm{Earth(8GHz)})}$ = 10$^{-23+7}$ = \mbox{10$^{-16}$ ergs$^{-1}$cm$^{-2}$}. Then, a channel power {value of} 
 \begin{eqnarray}
(\mathrm{Power})_{\mathrm{8Ghz}} = \mathrm{L_{k} \times PORTION}=10^{38} \mathrm{ergs}^{-1} \times \mathrm{PORTION}  \notag \\
=4\pi D^{2} \times I_{(\mathrm{Earth(8GHz)})}=4 \pi D^{2} \times 10^{-16}\mathrm{ergcm}^{-2}\mathrm{s}^{-1}  \notag \\
 = 4 \pi  \times (4 \times 10^{44} \mathrm{cm}^{2}) \times  10^{-16}\mathrm{ergcm}^{-2}\mathrm{s}^{-1}  \simeq 10^{30}\mathrm{ergs}^{-1}
\end{eqnarray}
 is obtained. Consequently, the value of PORTION can be estimated to be of the order of 10$^{-8}$. Therefore, our image includes (1/30) $\times$ (PORTION) $\times$ (L$_{\rm{k}}$). 


\section{Equipartition}\label{appb}

\label{equipartition}

The equipartition calculation follows. As shown above, jet kinetic power is 

\begin{equation}
 L_{\rm{k}}= P_{\rm{k}} = dE_{\rm{k}}/dt = (\Gamma-1) (N_{\rm{p}} m_{\rm{p}} N_{\mathrm{cell}} L^{2}_{\mathrm{cell}}) \beta c^{3} 
\end{equation}
where $\frac{dm}{dt}=\rho \frac{dV}{dt}=\rho A \frac{dx}{dt} =\rho A u$

Kinetic energy density is \cite{Reynoso2009}
\begin{equation}
\rho_{\rm{k}}=\frac{L_{\rm{k}}}{\pi R_{\rm{j}}^{2} u_{\rm{j}}}=\frac{L_{\rm{k}}}{A u}= \frac{(\Gamma-1) (N_{\rm{p}} m_{\rm{p}} N_{\mathrm{cell}} L^{2}_{\mathrm{cell}}) \beta c^{3}  }{A u}
,\end{equation}{}
At the nozzle, for the $\nu$-scale model, $\rho_{\rm{k}} \simeq$ 1~$\times$~10$^{8}$ ergcm$^{-3}$. For the heavier jet radio-scale model, $\rho_{\rm{k}} \simeq$ 1~$\times$~10$^{9}$ ergcm$^{-3}$ (3~$\times$~10$^{8}$ ergcm$^{-3}$ temporal average due to intermittent jet), and for the lighter jet radio-scale model, $\rho_{\rm{k}} \simeq 1~\times~$10${^7}$ ergcm$^{-3}$ (3~$\times$~10$^{6}$ ergcm$^{-3}$ average).

Also
\begin{equation}
B=\sqrt{8 \pi \rho_{\rm{B}}}
\end{equation}
and 
\begin{equation}
\rho_{\rm{B}}=B^{2}/8 \pi
\end{equation}
{For} both radio models, $\rho_{\rm{B}} \simeq$~10. For the $\nu$-scale model, $\rho_{\rm{B}} \simeq$~10$^{7}$. For equipartition, the kinetic and magnetic energy densities have to be equal to each other, $\rho_{\rm{k}}=\rho_{\rm{B}}$. Thus, both radio models have kinetic energy density higher than equipartition, whereas the $\nu$-scale model roughly fullfills the equipartition assumption. 




\section{RLOS Special Relativistic Imaging Code}\label{appc}

Additional properties of rlos imaging code are described here. 

\subsection{Theoretical Background for Imaging Code}

Refs. \cite{Hsiung1989, Hsiung1990} provide an early computerized attempt to reconstruct a relativistic image, through the eyes of an observer crossing a scene at high velocity. Ref.~\cite{Kraus2000} demonstrates the importance of the relativistic transform of brightness and color. When imaging a jet, these correspond to Doppler boosting and frequency shift, respectively. \cite{Kraus2000} discusses an object that moves at uniform speed across the field of view, but is visually large enough for the angle between velocity and line-of-sight to vary along the object. Applying the Lorentz transform changes brightness and color in a separate manner, for each point of the observed object.  Ref.~\cite{Weiskopf01} improves on such calculations, providing various methods for relativistic visualization, in both Special and General relativistic frameworks.

Ref.~\cite{Muller14b} calculate the visual appearance of wireframe relativistic objects, by mathematically inverting the course of light, from an image point to the emission event. They provide expressions that directly describe how a series of objects would look like, when moving at high speed, in front of a stationary observer. The efficiency of their method is then compared to the increased detail of a related ray-tracing project \cite{Muller14}. Ref.~\cite{Jarabo2015} image scenes with a fast observer traveling through their artificial environment. They also relate their simulations to actual imaging experiments, using the femto-photography technique~\cite{Velten13}. Furthermore, they introduce a number of additional details into their models, such as camera distortions from traveling at very high speed. Ref.~\cite{Zachary2016} present a framework, where the subject of relativistic imaging is explored, in an accessible manner.

\subsection{Time-Resolved Imaging}

\subsubsection{Accessing 4-Dimensional Data}

The finite nature of the speed of light crucially affects the appearance of a fast-moving object. Consequently, drawing a relativistic image of an astrophysical system, necessitates the availability of information regarding both  its spatial properties and its temporal evolution. In the present case, when executing the hydrocode, before running rlos, the temporal density of snapshots to be saved to disk at regular intervals, was suitably adjusted. The smaller those intervals are, the better the temporal resolution of hydrocode data are. A series of snapshots are then  loaded to RAM by rlos, which thereby requires more memory in order to run properly than the hydrocode itself does. Time is measured in simulation time units, which are read by PLUTO's attached pload.pro routine, which loads data into~rlos. 

The total time span available to an LOS (as measured in simulation time units, not merely in number of snapshots), $\Delta$t$_{\rm{LOS(total)}}$ = t$_{\rm{(last-shot)}} -$ t$_{\rm{(first-shot)}}$ ({\label{tshot}not to be} {confused with interval $\Delta$t$_{\rm{shot}}$ between \emph{successive} snapshots}) should be preset to be larger than light crossing time of the model system, for the selected LOS angle settings. Documenting model jet evolution generally requires hydrocode data saves to be rather dense in time, especially for fast-changing flows. On the other hand, a lower temporal resolution  probably suffices for a steadier, slower-paced flow.

\subsubsection{Traversing  4D Arrays}

\paragraph{Introduction}

\begin{figure}[H]\vspace{-12pt}
\includegraphics[width=10.5 cm]{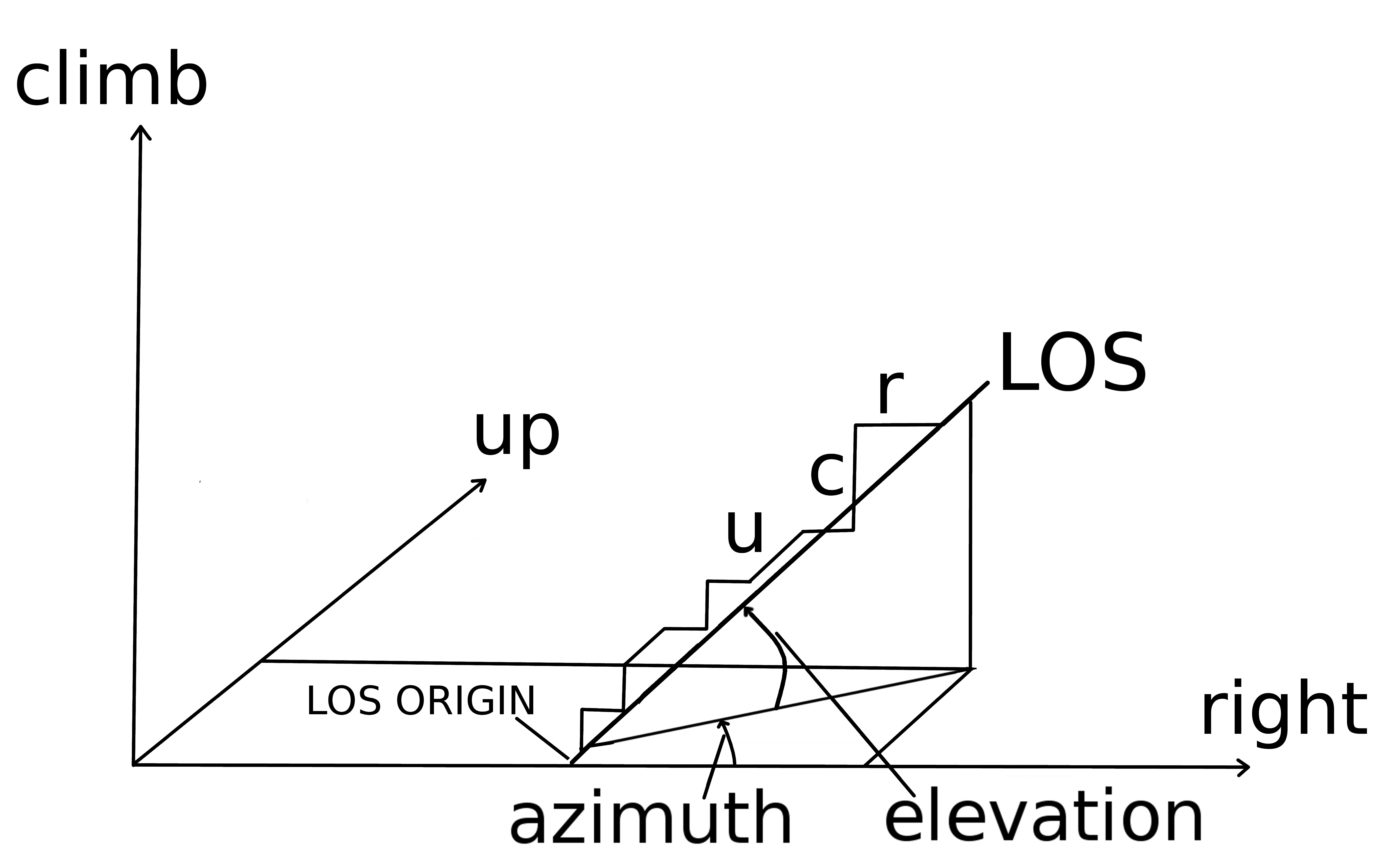}
\caption{Schematic of the spatial propagation of a line-of-sight (LOS) through a 3D Cartesian computational grid. In the discrete grid, according to the design of the algorithm, there are 3 available directions to be taken at each step along the LOS: \emph{right}, \emph{up}, and \emph{climb}. These correspond to x, y, and z, respectively. During propagation, the LOS tries to follow its given direction, as defined by the two angles of azimuth and elevation. More specifically, every two steps, a decision is first made on azimuth, either right or up. Then, for elevation, it is either climb or another azimuth decision. Along the LOS, horizontal steps point to the ‘right’ direction. Diagonal steps represent going ‘up’, while vertical ones constitute ‘climb’ steps.}
\label{losdraw}
\end{figure}

\begin{figure}[H]
\includegraphics[width=9 cm]{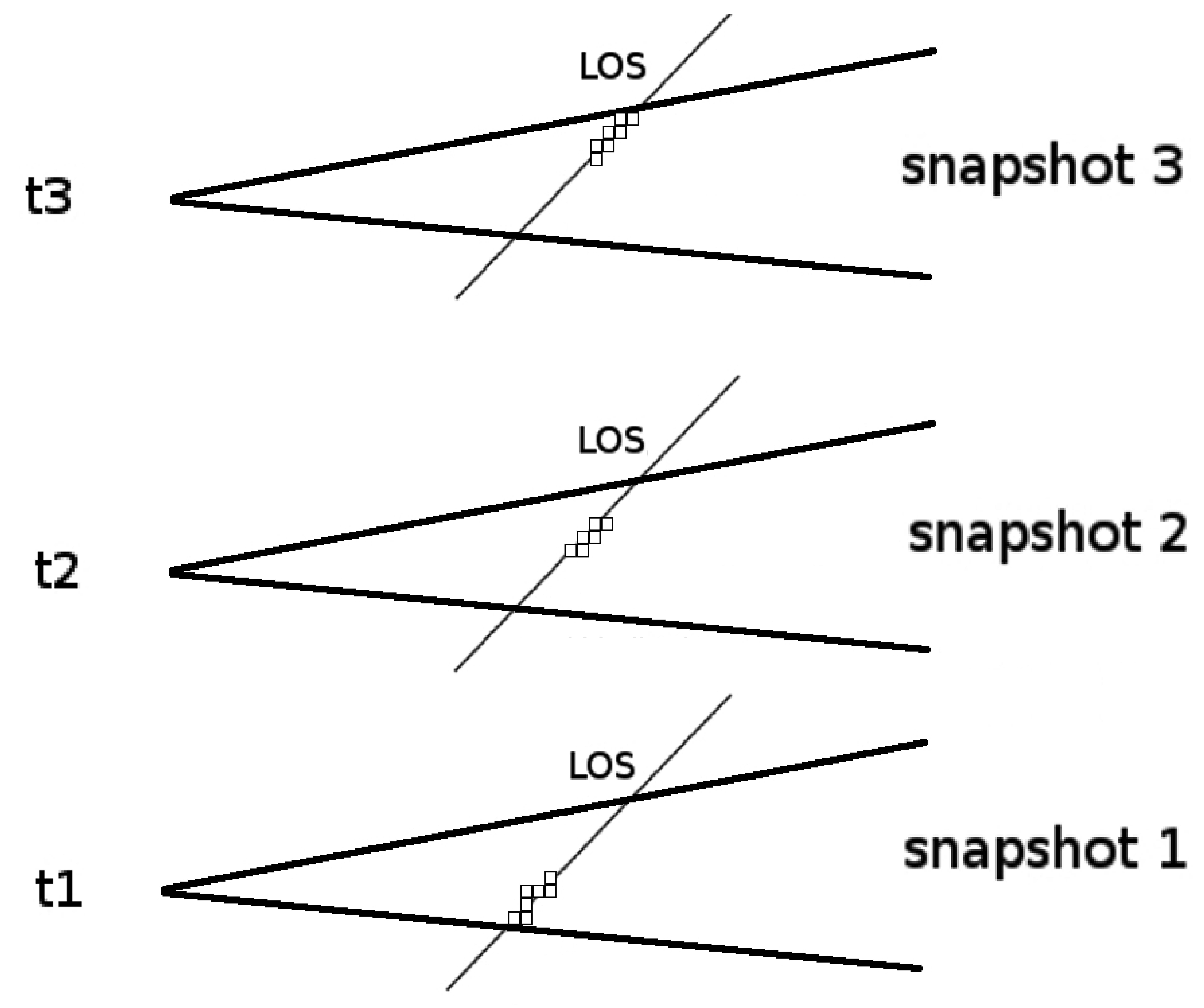}
\caption{{Three successive instants} 
 of a line-of-sight traversing a jet. At regular intervals, we jump to a new 3D slice of a 4D spacetime array, obtaining a discrete approximation of the time continuum in the form of hydrocode snapshots. The calculation may proceed either forward in time (from bottom to top) or backwards in time (from top to bottom).}
\label{4dfigure}
\end{figure}

\begin{figure}[H]
\includegraphics[width=10.5 cm]{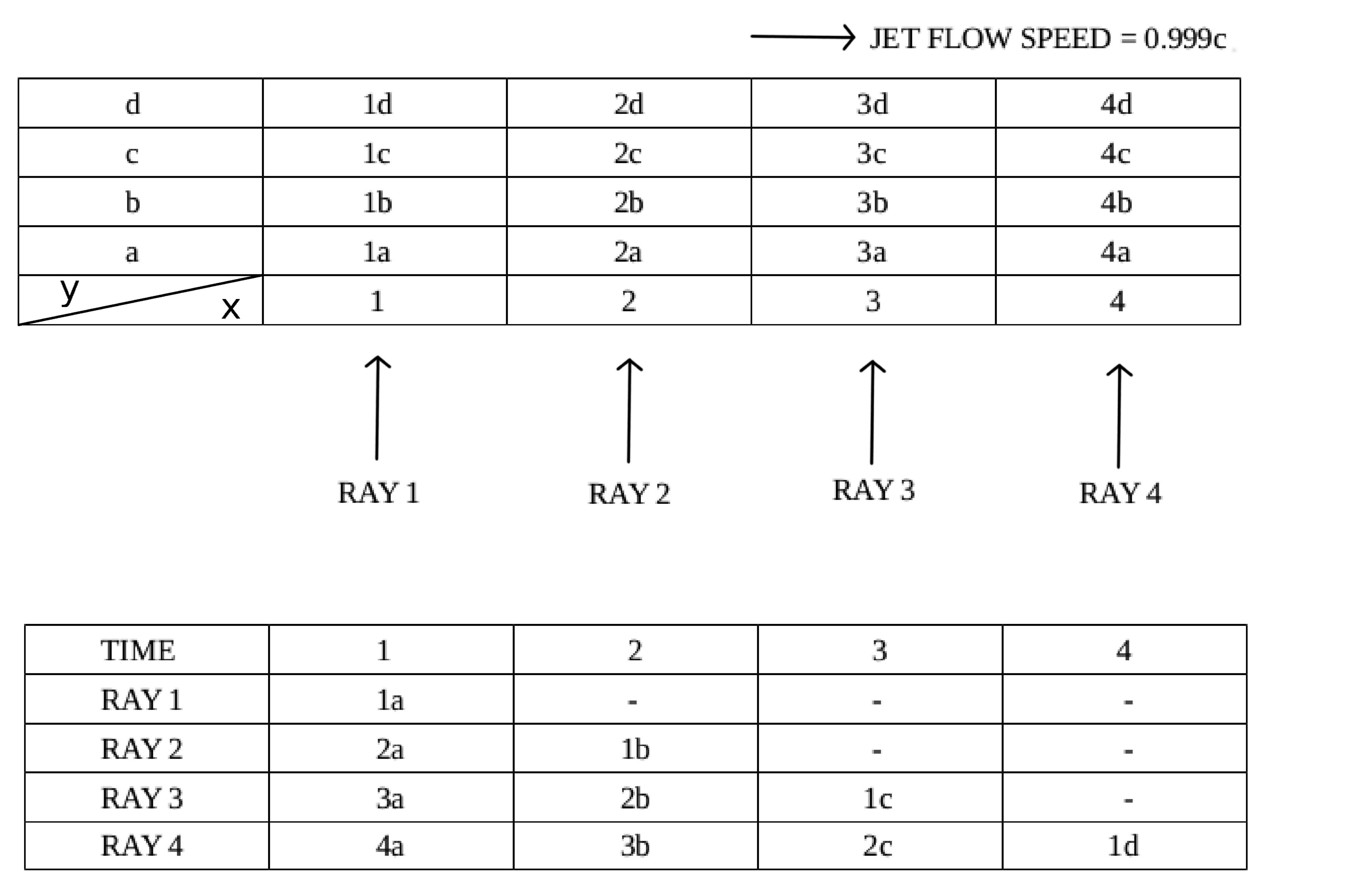}
\caption{Simultaneous advance in two-dimensional space and forward in time of a few lines-of-sight. \textbf{Top half} depicts the spatial situation at $t=1$. Sixteen jet matter portions  occupy this mini 4 by 4 grid. Each piece of matter is named after its position at t = 1 and retains that name as it moves along. \textbf{Bottom half} shows how the situation evolves as time marches on, with light rays meeting different jet segments that cross their path. A dash means a light ray meeting jet matter other than the above or nothing at all.}
\label{pinakes}
\end{figure}

A series of hydrocode snapshots are loaded to RAM, populating the elements of 4-dimensional (4D) arrays. Let us examine the case of backwards in time imaging calculation. From a temporal point of view, we begin from the simulation time corresponding to the last of the loaded snapshots, called shotmax. From a spatial point of view, we start at a point of the imaging plane, which is a side of the computational box (Figure \ref{3Dgeometry}). As the calculation advances, in 3D space along the LOS being drawn (Figure \ref{losdraw}), the algorithm keeps checking whether to jump to a previous temporal slice while staying on target in 3D (Figure \ref{4dfigure}). Consequently, the LOS advances back in time through data by accessing different instants from the 4D data arrays. As a test case, the LOS may also be set to advance forward in time, beginning from the time tag of shotmin, the first of the loaded snapshots (Figures \ref{4dfigure} and \ref{pinakes}). 

\paragraph{Time-Resolved Imaging Calculations}

For every LOS, there is a point of origin (POO), located on the imaging side of the computational grid (Figure \ref{3Dgeometry}). That point, addressed in rlos code as (nx10, ny10, nz10) and here as O$^{\prime}$, is the beginning of the LOS's axes x$^{\prime}$, y$^{\prime}$, z$^{\prime}$, parallel to x, y, and z, respectively. A 2D loop covers the imaging surface, the POO successively locating itself at each of the latter's  points.

As we progress along an LOS, a record is kept of where we are, in 3D space. This record comprises the LOS's own integer coordinates, rc, uc, and cc, measured in cells, starting from its POO. The above symbols stand for right-current, up-current, and climb-current, representing the current LOS advance in the x$^{\prime}$, y$^{\prime}$, and z$^{\prime}$ axes, respectively (\mbox{Figures \ref{3Dgeometry} and \ref{losdraw}}). The current ray position is then (nx10 + rc, ny10 + uc, nz10 + cc).

A timer variable, curtime (standing for current LOS time), is introduced for each LOS, recording the duration of insofar ray travel along the LOS. The aforementioned timer is preset at the beginning of each LOS to the hydrocode time of the first loaded data snapshot (forward in time integration), or of the last snapshot (back in time integration). For backwards in time ray-tracing, the above duration is subtracted each time from t(shotmax). 



\vspace{-10pt}

We then proceed to calculate the current length of the LOS
\vspace{-15pt}

\begin{eqnarray}
\label{loslength}
 \nonumber 
l_{\rm{los(current)}}=[ (\rm{dlr}*(\rm{nx1current}- \rm{nx10}))^{2}+ \notag\\ 
(\rm{dlu}*(\rm{ny1current}- \rm{ny10}))^{2}+(\rm{dlc}*(\rm{nz1current}- \rm{nz10}))^{2} ]^{1/2}
\end{eqnarray}
%
where  LOS length is measured in cell length units and
\begin{equation}
\label{nx1current}
\rm{nx1current}=\rm{nx10}+ \rm{rc}, \,\, \rm{ny1current}= \rm{ny10}+ \rm{uc}, \,\, \rm{nz1current}= \rm{nz10}+ \rm{cc}
\end{equation}
along the x, y, and z directions, dlc, dlu, and dlr are the respective \emph{normalized} hydrocode Cartesian cell lengths. Their values are usually unity or close to unity, as set in the hydrocode by the user, and rlos requires them fixed, meaning only homogeneous grids are currently supported. Furthermore, if the hydrocode grid is read by pload at a reduced resolution,  rlos cell sizes are automatically adjusted accordingly.

We can finally write
\begin{equation}
\label{loslength2}
l_{\rm{los(current)}}=[ ((\rm{dlr}* \rm{rc})^{2})+((\rm{dlu}* \rm{uc})^{2})+((\rm{dlc}* \rm{cc})^{2}) ]^{1/2}
\end{equation}
{We} then proceed to calculate curtime, the current hydrosimulation time of the light ray along the LOS.
For forward in time LOS advance
\begin{equation}
\label{curtime}
\rm{curtime}=\rm{l}_{\rm{los(current)}}/\rm{c}_{\rm{light}}+t_{\rm{shotmin}}.
\end{equation}
while for back in time ray-tracing
\begin{equation}
\label{curtime}
\rm{curtime}=+ \rm{t}_{\rm{shotmax}}- \rm{l}_{\rm{los(current)}}/\rm{c}_{\rm{light}}.
\end{equation}
t$_{\rm{shotmin}}$ is the timestamp of the first loaded snapshot, t$_{\rm{shotmax}}$ the one of the last snapshot loaded, and clight is the speed of light in cells per simulation second.

When curtime crosses a new snapshot's time tag, the algorithm switches to drawing the LOS through the 3D volume of the new snapshot (Figure \ref{4dfigure}). We keep moving along the same LOS in 3D space, but we   switch to a new time instant in the data. The above temporal shift is repeated as many times as required by the relevant criterion along the LOS until the spatial end of the LOS.

\subsubsection{Aiming at the Line-of-Sight}
\label{aimingthelos}

The direction of an LOS in 3D space is defined by the two angles of azimuth (angle 1) and elevation (angle 2) (Figure \ref{3Dgeometry}), where the plane of angle 1 is the x$^{\prime}$y$^{\prime}$ plane, parallel to xy. 

\begin{figure}[H]\vspace{-13pt}
\includegraphics[width=10.5 cm]{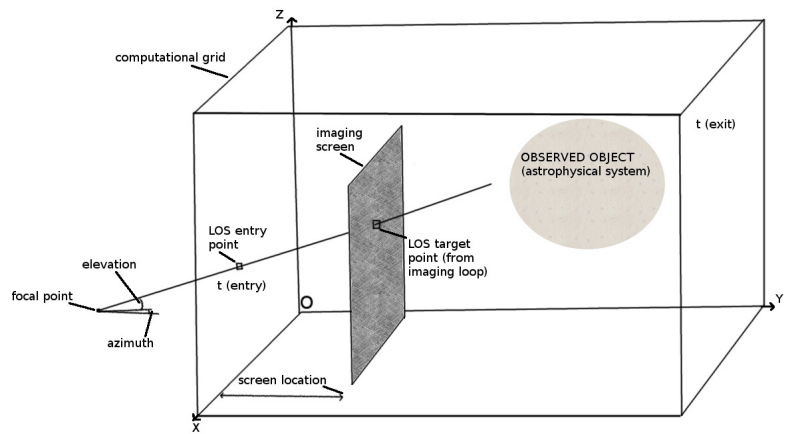}
\caption{Geometry of focused-beam imaging in rlos, for the case of forward in time calculation. A fiducial screen is where the image is formed.}
\label{focused_beam_imaging_geometry}
\end{figure}

LOS's may be either parallel to each other, or focused. With a focused beam, the image is formed on a adjustable size fiducial screen located at a user-defined position between the focal point and the model system (Figure \ref{focused_beam_imaging_geometry}).  

\vspace{-6pt}

\begin{figure}[H]
\includegraphics[width=10.5 cm]{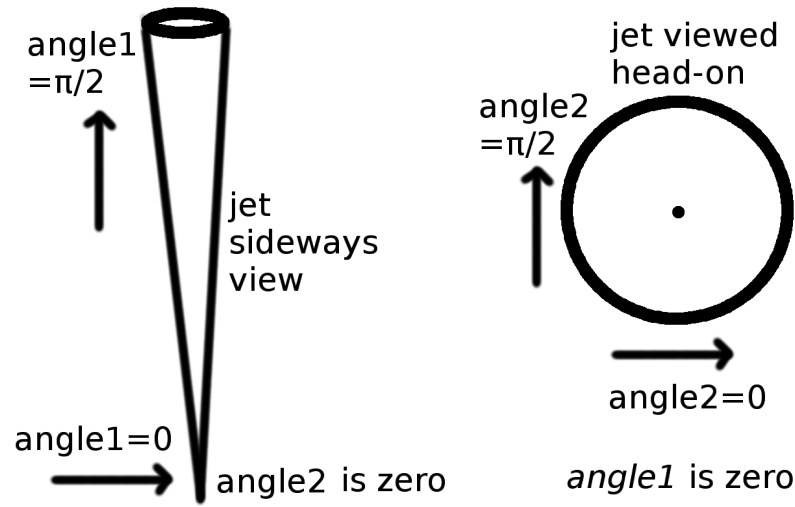}
\caption{Geometric arrangement with regard to the viewing angles in the model for the special cases of angle2 = 0 (\textbf{left}) and angle1 = 0 (\textbf{right}). For each subcase, the arrow shows the LOS direction, which is different than the reader's direction of view.}
\label{angles}
\end{figure}

For a jet parallel to the y axis, the angle between local jet matter velocity $\vec{u}$ and  LOS, losu = $(\widehat{\vec{\rm{LOS}},\vec{u}})$, is generally small when angle 1 (xy azimuth) approaches 90 degrees and vice versa (Figure \ref{angles}). As is well-known \cite{Hughes1991}, the angle losu affects the relativistic emission calculations. {Individual jet elements may still move in directions different than the main jet axis, as part of a dynamic flow.}

For a jet along the y axis, the plane of angle 2 (elevation) is largely perpendicular to the jet when azimuth (angle 1) is zero, while it is roughly parallel to the jet when azimuth (angle~1) is 90 degrees. Usually, the jet bears an approximate cylindrical symmetry, and this has an interesting effect on the sensitivity of the synthetic image to the viewing angles (Figure \ref{angles}). More specifically, for a small azimuth (angle 1), if we vary elevation (angle 2), we indeed rotate the viewing point around the jet axis, thus producing similar intensities, thanks to the approximate cylindrical symmetry of the jet. Thus, for a jet moving along the y~axis, the smaller azimuth (angle 1) is, the less difference varying elevation \mbox{(angle 2) makes.}

On the other hand, for azimuth (angle 1) nearing $\pi$/2, varying elevation (angle 2) rotates the view within a plane approximately parallel to the jet, resulting to considerable differences in the image (no symmetry involved this time). Consequently, the larger azimuth (angle 1) is, the stronger the effect, on the synthetic image, from changing elevation (angle 2).


\subsection{Relativistic Effects-Doppler Boosting}
\label{Dboosting}

\subsubsection{General}

The main effects of the Lorentz/Poincar\'e transform on the emission from a relativistic object \cite{Weiskopf01}, specifically applied to an astrophysical jet, are relativistic aberration, time dilation, and frequency shift \cite{Cawthorne1991, Hughes1991, Sparks92, Laing2002}.

This area refers to E/M emission, but not to particle emission. Jet spectrum is given by any suitable form inserted into the model, including the spectrum resulting from synchrotron emission and absorption coefficients. Earth frame jet spectrum S$_{\rm{obs}}$ can be expressed \cite{Hughes1991, Cawthorne1991} as 
\begin{equation}
S_{\rm{obs}} = S_{\rm{jet}} D^{3+\alpha} 
\label{totalboosting}
\end{equation}
where $\alpha$ is the spectral index and D is the Doppler factor. Exponent (3 + $\alpha$) in the above can be broken down into different contributions from separate effects. Two units come from the aberration of light, one from the relativistic dilation of time and $\alpha$ from the effect of frequency shift, while for a continuous optically thin jet, a D factor is lost \cite{Cawthorne1991}.


\paragraph{Aberration-Searchlight Effect} 

Relativistic aberration changes the perceived direction of light (there is no ray curving in special relativity) when transforming between the jet frame and the earth frame, ‘tilting rays’ emanating from the jet, generally towards its head area. 

 
At an individual cell level, cell emission along a ray {\emph{within}} 
 the cell's boost cone is then accordingly reinforced; if {\emph{outside}} the cone, it is weakened. Depending on local velocity value and direction, neighbouring cells may have totally different boost cones.


\paragraph{Time Dilation}

Time dilation contributes one D factor to the emission result. Again, this refers to E/M~radiation.

\paragraph{Frequency Shift}

E/M radiation emitted at a given frequency, from fast-moving jet matter, is taken to be Doppler shifted in frequency
\begin{equation}
\label{ncalc}
f_{\rm{obs}} = f_{\rm{calc}}D
\end{equation}
where $f_{\rm{obs}}$ is the observed frequency, and $f_{\rm{calc}}$ is the frequency used in emission calculations performed in the jet frame of reference \cite{Hughes1991}. 

For D $\geq$ 1, emission is calculated at a frequency lower than the observed, resulting in higher intensity, when the employed spectrum decreases with frequency. For radio emission, we therefore calculate emission and absorption at the $f_{\rm{calc}}$ = $f_{\rm{obs}}$/D. Each computational cell, in general, has its own Doppler factor D, and therefore its own frequency~shift.

\subsubsection{Lorentz Factor}









The Lorentz factor for a hydrocode cell is \cite{Hughes1991}

\begin{equation}
\label{gammalorentz}
\Gamma_{\rm{Lorentz}} = \frac{1}{\sqrt{1-u^{2}}}
\end{equation}
where
\begin{equation}
u=\sqrt{u^{2}_{x} + u^{2}_{y} +u^{2}_{z}} \leq 1
\label{speedcomponents}
\end{equation}
is the value of  local velocity $\vec{u}=(u_{x},u_{y},u_{z})$, in units of the speed of light.




\subsubsection{Doppler Factor Calculation}
\label{Appendix_coslosu_calc}
Jet radiation is either boosted or deboosted, depending on the angle ‘losu’ 
between the direction of the LOS and $\vec{u}$. The higher the jet speed is, the narrower and stronger the cell boost cones are, around the direction of local velocity. On the other hand, outside cell boost cones, deboosting occurs, that is to say, the higher the velocity is, the weaker the signal becomes. D equals
\begin{equation}
D=\frac{\sqrt{1-u^{2}}}{(1-u* \cos(\rm{losu}))}
\label{doppler_factor}
\end{equation}

For the above, the angle between the LOS and the local velocity vector is required at every point of the computational space. 

(As a note, for particles, their distribution is transformed to the Earth frame, as shown in \cite{TR11}). 

The cosine of angle losu is calculated in the following manner:

\textls[-15]{Let us define a fiducial unitary LOS vector  $\vec{(\rm{LOS})}=(\rm{lx_{1}, lx_{2}, lx_{3}})$, with $(\rm{LOS})=\sqrt{  lx_{1}^{2} + lx_{2}^{2} + lx_{3}^{2}  }=1$.}
In the following, $\phi_{1}$ and $\phi_{2}$ represent azimuth and elevation  angles 1 and 2, respectively.
\begin{equation}
\mathrm{lx} _{1}= \cos(\phi_{1}) \cos(\phi_{2}), \,\, \mathrm{lx}_{2}= \sin(\phi_{1}) \cos(\phi_{2}), \,\, \mathrm{lx}_{3}= \sin(\phi_{2}) 
\label{directions}
\end{equation}
\begin{equation}
\vec{\rm{LOS}} * \vec{u}=(\rm{LOS}) u \cos(\widehat{\vec{\rm{LOS}},\vec{u}})=lx_{1} u_{x}+lx_{2} u_{y}+lx_{3} u_{z}
\label{coslosu1}
\end{equation}
{Therefore}, we have ((LOS) = 1)
\begin{equation}
\cos(\widehat{\vec{\rm{LOS}},\vec{u}})=\frac{ \rm{ lx_{1}u_{x}+lx_{2}u_{y}+lx_{3}u_{z} } }{  (\rm{LOS}) u  }=\frac{ \rm{ lx_{1}u_{x}+lx_{2}u_{y}+lx_{3}u_{z} } }{  \sqrt{ \rm{ (  u_{x}^{2} + u_{y}^{2} + u_{z}^{2}  )  } } }
\label{coslosueqn}
\end{equation}
{For} back in time ray-tracing, a minus sign is introduced to the above equation. Furthermore, a minuscule number is added to the denominator of Equation (\ref{coslosueqn}), in case \mbox{u = 0}. The above calculation allows the assignment of a Doppler boosting factor through \mbox{Equations~(\ref{speedcomponents}), (\ref{doppler_factor}) and (\ref{coslosueqn})} to each discrete emission event along a line-of-sight.

\subsubsection{CoslosB Calculation}
\label{Appendix_coslosb_calc}

The calculation of the angle losb between the magnetic field and the LOS is performed, in rlos, in exactly the same way as for losu, above. Only this time the vector of velocity is replaced by the magnetic field one.

\paragraph{Alternative Frequency Shift}

rlos may include different emission dependencies on frequency, where we calculate intensity at f$_{\rm{calc}}$ and observe that at f$_{\rm{obs}}$. 



\subsection{Testing Parameters}
\label{testingparameters}

Certain parameters, that facilitate testing rlos, are presented here.

\subsubsection{The Clight Parameter}

Let us consider a 4D array, comprising a succession of hydrocode snapshots. The LOS traversing those data, moves at a speed of clight cells per time unit. When we artificially adjust clight to a lower value \cite{KTS13, Zachary2016}, then the algorithm jumps to a new snapshot after spatially advancing through fewer cells. A slower LOS advances farther in time while crossing a given distance through the jet, allowing for a detailed study of the time-jumping algorithm. On the other hand, setting clight to a very high value leads to a single shot image, as we never advance to a further temporal slice. 

A representative hydrocode scaling is the following (for neutrinos)
\begin{align}
\label{pluto_units}
L_{\mathrm{sim}} = 10^{10} \mathrm{cm}, \,\, u_{\rm{sim}} = 3 \times 10^{10} \frac{\mathrm{cm}}{\mathrm{s}}, \rho_{\rm{sim}} = 1.67 \times 10^{-24} \frac{\mathrm{g}}{\mathrm{cm}^{3}} \,\, t_{\rm{sim}} = \frac{L_{\rm{sim}}}{u_{\rm{sim}}} = \frac{1}{3} \mathrm{s}
\end{align}
where $t_{\rm{sim}}$ is the hydrocode time unit, u$_{\rm{sim}}$ is the speed of light and $L_{\rm{sim}}$ is the hydrocode length unit. When preparing the hydrocode run, the time span, in simulation seconds, between data snapshots, should optimally be set, to l$_{\rm{LOS}}$/(n*clight). l is the LOS length, in \emph{cells} and n is the desired number of snapshots to cover the imaged timespan. If we employ the parameter sfactor, pload's shrink factor, cells are enlarged and the calculated value of clight shrinks accordingly (sfactor regrids the hydrodata to a coarser grid). Overall accuracy then suffers somewhat, and regridding to a lower resolution should be used only as a preview. 

Altering clight only affects the light ray speed, not the speed of matter. Consequently, overriding clight does not affect the relativistic emission calculations (like tweakspeed does, Appendix \ref{tweakspeed}). An altered clight is merely an artifice, introduced in post processing, in order to explore the effect of using more, or less,  temporal slices in the final image.

\begin{figure}[H]
\includegraphics[width=10.5 cm]{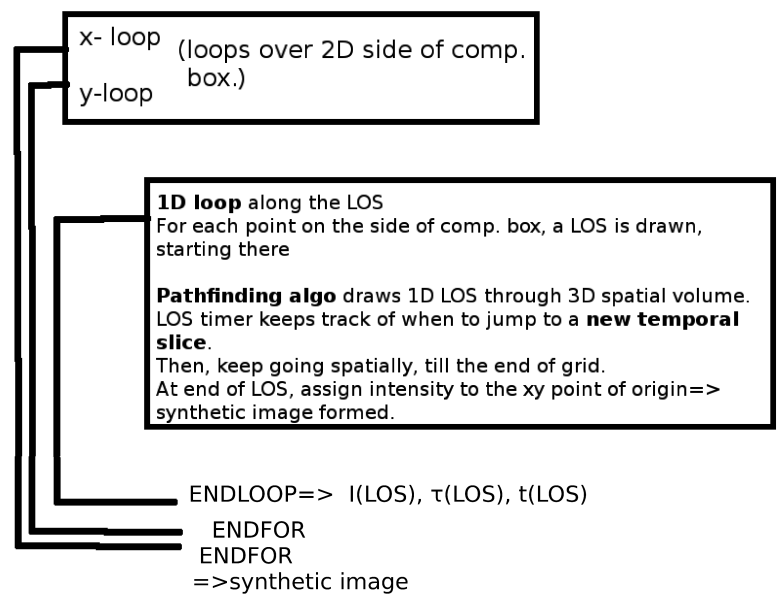}
\caption{A simplified flow diagram depicting the basic logical structure of rlos imaging code, for the case of rays parallel to each other. The synthetic image's xy loops here correspond to either the yz or the xz side plane of the computational box.}
\label{flow_diagram1}
\end{figure}


\subsubsection{The FS switch}

After the hydrodata are loaded, a global operation calculates, for each cell, a jet frame frequency $f_{\rm{calc}}$: $f_{\rm{calc}}=f_{\rm{obs}}/D$, where $f_{\rm{obs}}$ is the observing frequency and D is the local Doppler factor of the cell. The frequency shift (FS) switch selects between using the local f$_{\rm{calc}}$ or the global f$_{\rm{obs}}$ in the emission calculations.  The FS facility allows for a direct user-defined emission dependence on frequency to be introduced (\cite{Cawthorne1991} pg. 199: Simplified synchrotron jet), in the form of of a function $S_{\rm{obs}}=S_{\rm{obs(f)}}$. 

\subsubsection{The DB Switch}

The DB switch offers the option of using the Doppler boosting effect.

\subsubsection{The Speed Tweak Parameter}
\label{tweakspeed}

A test is introduced, whereby matter velocity is multiplied, on a global scale, by a ‘speed tweak’ factor. This offers a quick way to observe the impact, on the synthetic image, of altering the hydrodynamic speed in post-processing, for the same simulation run. The natural value of tweakspeed is 1. At low tweak speed factors (less than 1) the effects, on the final image, of both DB and FS, are reduced, and vice versa. The maximum for tweakspeed is c/u$_{(\rm{max)}}$, above which velocites higher than c are artificially created in the grid.

\subsection{rlos210}

rlos version 2.10 includes a unified, functionalized, modular approach. The XZ and YZ versions were merged for both focused beam and parallel LOSs. Code was reorganised, and a series of tests are included.

\subsection{rlos210 Commentary Transcript}



\textls[-15]{This is the latest version of rlos. Version 2 is a major upgrade of original rlos code. This time, the programme was broken up into procedures and functions with a modular~structure.  }

The programme allows the user to select which case to simulate, through an external parameter file. There is a unified approach, where the same modules operate on different geometries, through parameterization.  

The user may select the values of the parameters of rlos version 1, and fully employ them. As mentioned above, there is no more a different version of rlos for XZ and YZ plane image formation. Now, there is one version of the code for both cases. Furthermore, for each of those cases the user may select either radiograph or camera obscura imaging~technique. 

The radiograph setup has all lines of sight parallel to each other, just like rlos v.1. This means the film (fiducial imaging screen) is the size of the scene (grid), like an X-ray medical image. The latter type of image shows clearly the various details of the system.   

On the other hand, camera obscura, or focused beam, has a focal point where the eye of the fiducial observer is located. 
The imaging screen, in camera obscura, is of varied size: It may be equal, or smaller to the grid slice, at a given point along either the x or y axis depending on YZ or XZ imaging plane case.
At the moment, the fiducial imaging screen must be parallel to the corresponding side of the grid, i.e., either XZ or YZ. 
Screen location on-axis may vary within the grid. The smaller the screen is, the smaller the image.

The focal point may reside either on the side of the grid or outside the grid but within the limits of the projection of the XZ or YZ plane. It may have a negative or zero axis position, but its two planar coordinates must be smaller than the grid size.

Direction angles are no longer necessarily constant throughout the calculation: for the focused-beam case, each LOS is drawn with a different set of azimuth (phi1) and elevation (phi2) angles. Angles are calculated using the lines that connect the focal point and the imaging screen point, which is the target point for the LOS. 

The LOS then begins from the focal point if it resides on the grid side or from the LOS entry point, calculated 
suitably. From then on, it advances using aiming algorithms, trying to pass through the targeted screen point. It normally obtains the target or closely misses it. In general, the higher the resolution is, the better the accuracy in this respect.

\textls[-5]{For GR pseudo-Newtonian simulations, a logical next step is to introduce D(phi1), D(phi2), i.e., alter angles along an LOS from cell to cell according to the effect of the~potential.} 

Then, innermost jet workings may be imaged, if the hydrocode can employ influence from a black hole.


\subsubsection{Back in Time Integration along the LOS}
\label{back_in_time}

In this version, calculations may be performed either ahead in time or backwards in time from a selected time instant (tpicked) backwards. For camera obscura, back in time is generally the correct way to proceed. For radiographs, ahead in time also works fine, assuming a suitable fiducial setup of the jet system vs the observer. 

Tpicked is only employed when back in time switch is activated in the external parameter file. tpicked must be generally towards the end of the preselected range of dump files, or timeshots, to be loaded to RAM. At the moment, it is set equal to tmax, for convenience. A sufficient backwards time range must be provided for the LOS, in order to travel back through time without reaching the beginning of the grid. Otherwise, code cannot finish integration along the LOS. 
When testing, the facility of altering light speed, clight, may be used to study this effect. 

Pathfinding algorithms were  upgraded for this version. For each combination of XZ or YZ and radiograph or camera obscura, a certain set of such pathfinders were employed. 





\section{Neutrino Emission Calculations}\label{appd}

\label{neutrinos}

{The presentation in this section} 
 is based on the corresponding section of \cite{Smponias_2021}.
 
\subsection{Proton Energy Loss}

\begin{figure}[H]
\includegraphics[width=10.5 cm]{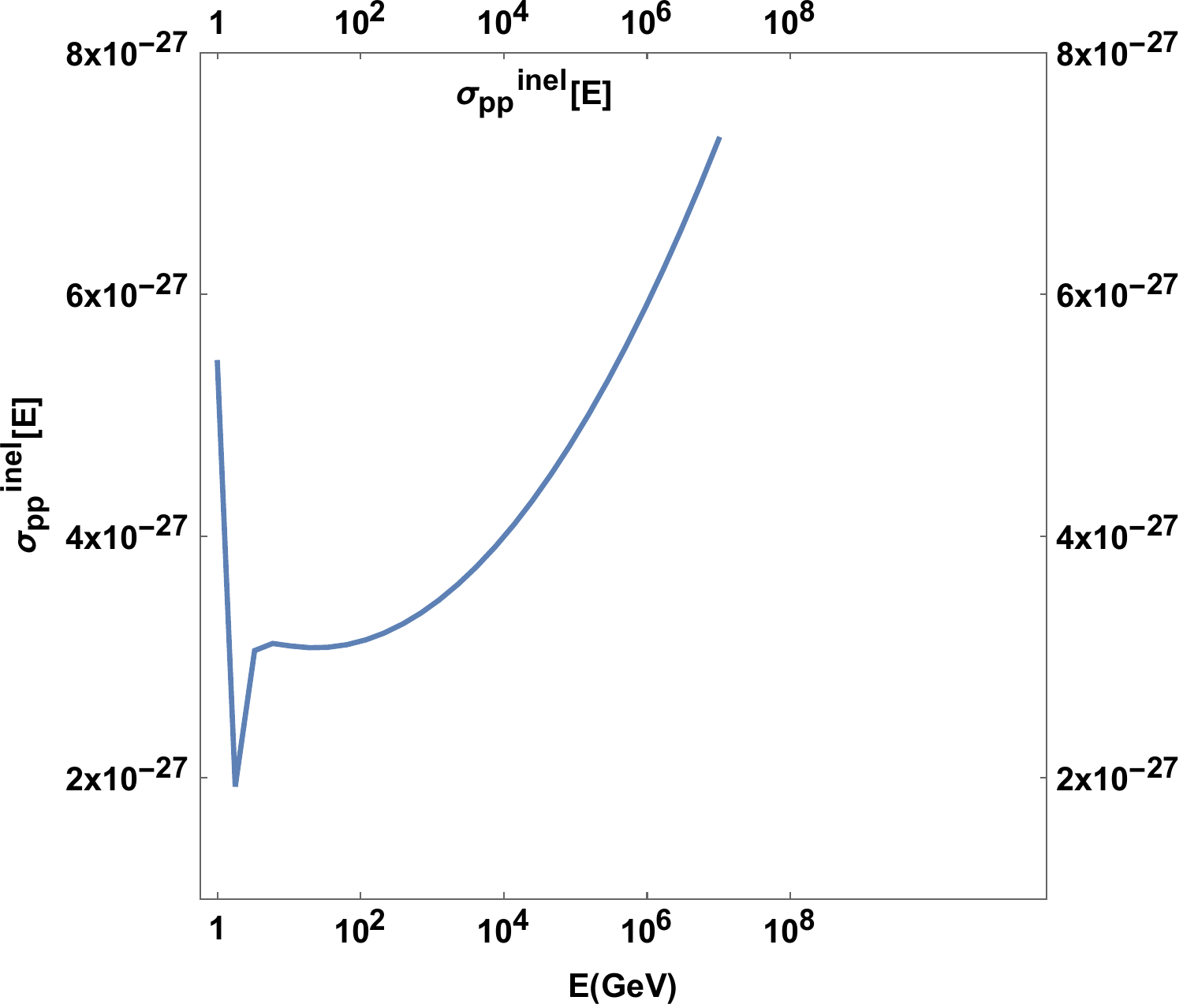}
\caption{{Inelastic proton--proton} 
 collision standard plotted with energy. It demonstrates  rather small variation (linear vertical scale) of its value over a large energy range (logarithmic horizontal scale), covering and exceeding the energy span required for the calculations that follow later in this~paper.} 
\label{sigma_inel_pp}
\end{figure}
\begin{figure}[H]\vspace{-23pt}
\includegraphics[width=10.5 cm]{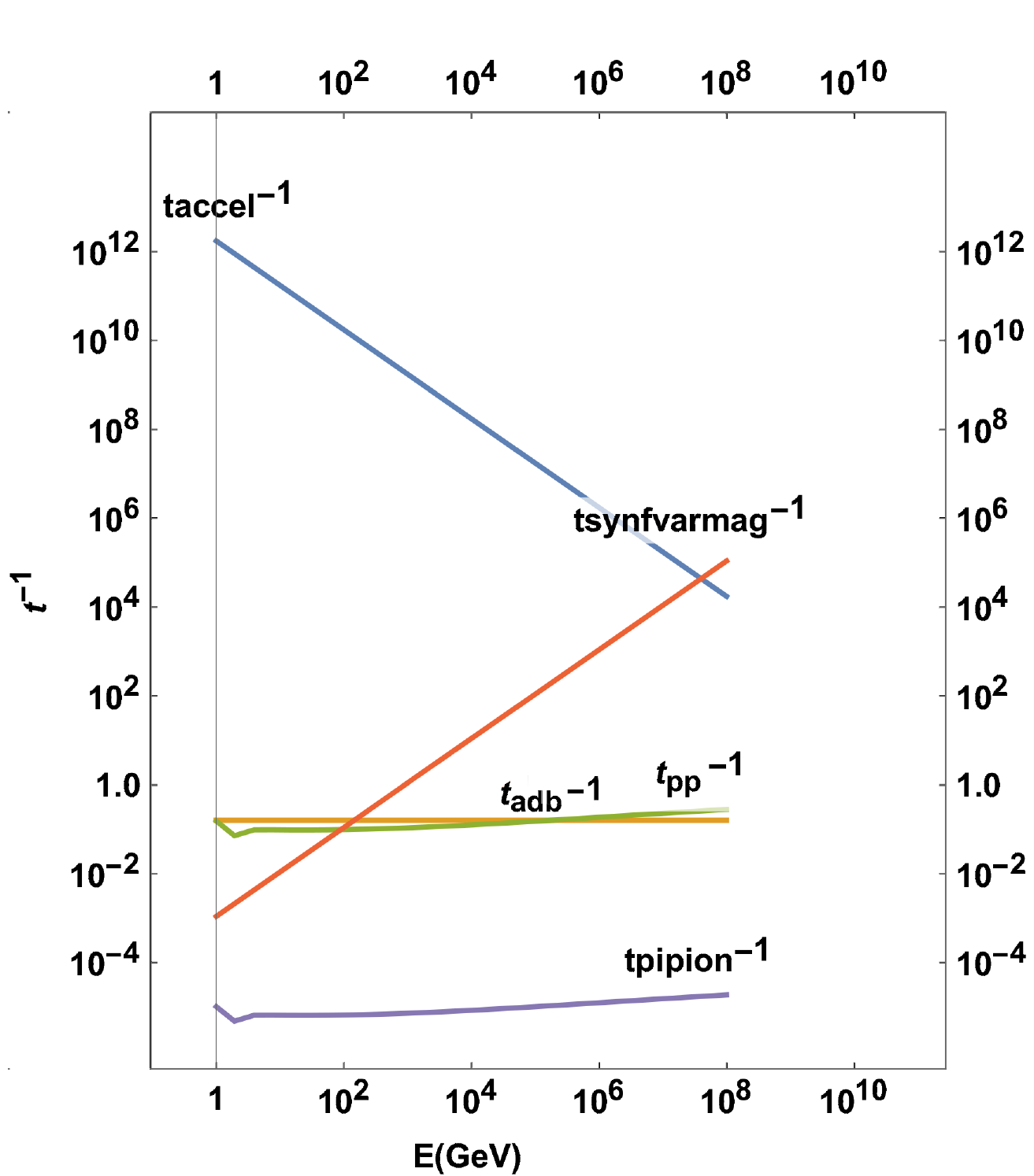}
\caption{{Non-thermal proton distribution energy} 
 loss time scales, for various loss mechanisms, drawn with energy in GeV. t$_{\mathrm{accel}}$ is the proton acceleration time scale at shocks. t$_{\mathrm{synfvarmag}}$ stands for the synchrotron mechanism loss time scale, using a magnetic field that varies from point to point within the jet. t$_{\mathrm{adb}}$ is the adiabatic loss time scale, t$_{\mathrm{pp}}$ is the (hot--cold) proton--proton collision timescale. t$_{\mathrm{pipion}}$ stands for the pion decay timescale t$_{\pi}$.}
\label{tloss_many}
\end{figure}
\begin{figure}[H]
\includegraphics[width=10.5 cm]{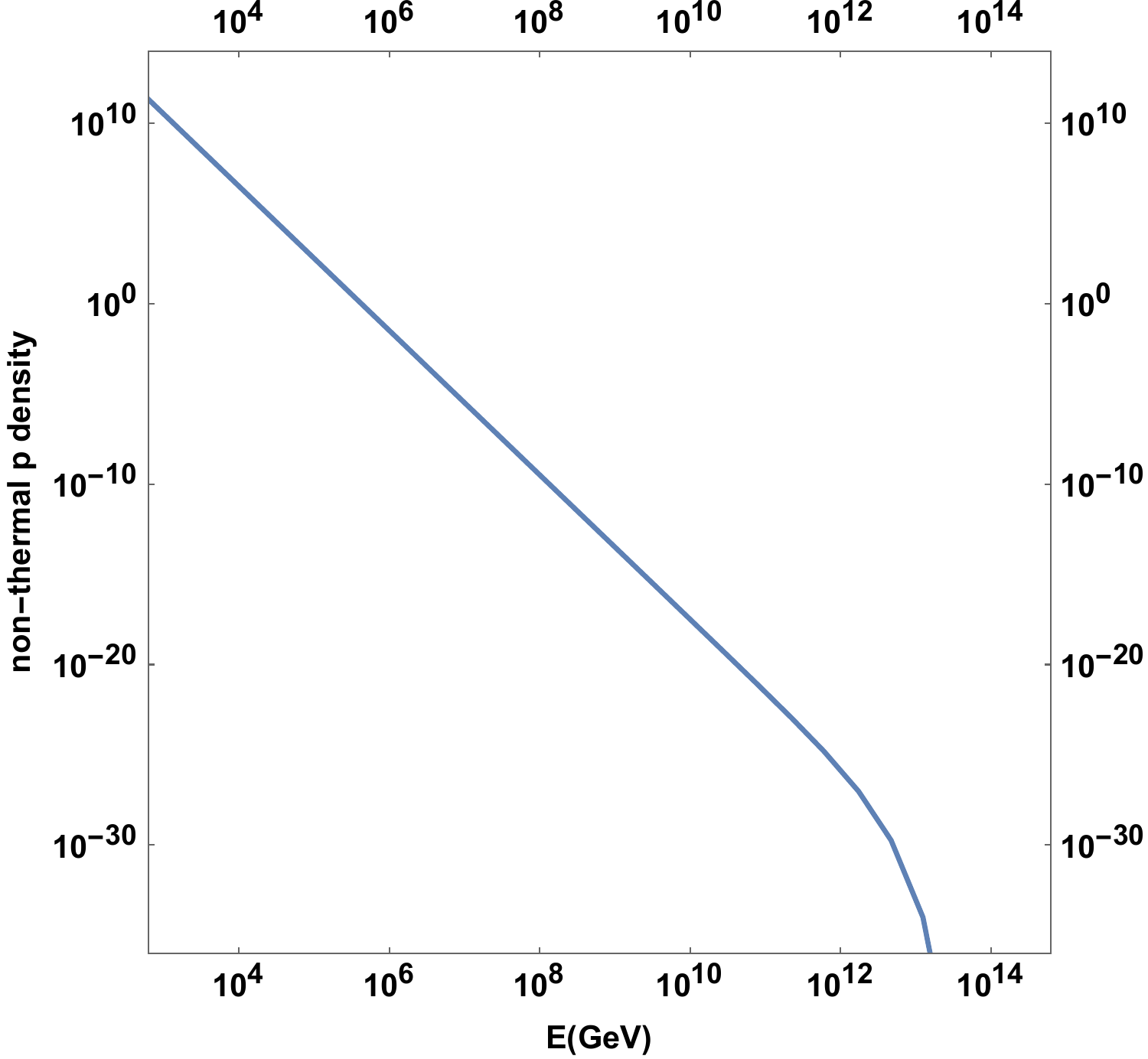}
\caption{Density of nonthermal protons in the jet using a high-energy cutoff feature plotted with~energy.}
\label{fast-p-density}
\end{figure}
\begin{figure}[H]\vspace{-6pt}
\includegraphics[width=10.5 cm]{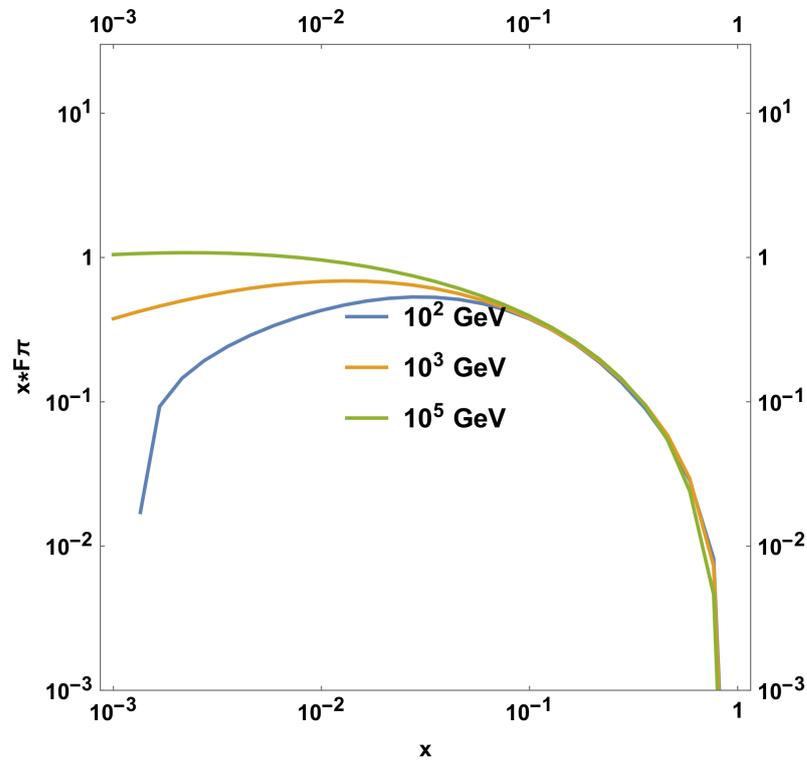}
\caption{  F$_{\pi}^{(\mathrm{pp})}\left( x,\frac{E}{x} \right)$ function, Equation (\ref{Ffunction}), corresponding to the pion spectrum emerging from a single (hot--cold) proton collision,  multiplied by x=$\frac{E_{\pi}}{E{\mathrm{p} }}$. Calculated at three different energies of the hot proton.}
\label{Fpp_plot}
\end{figure}
\begin{figure}[H]
\includegraphics[width=10.5 cm]{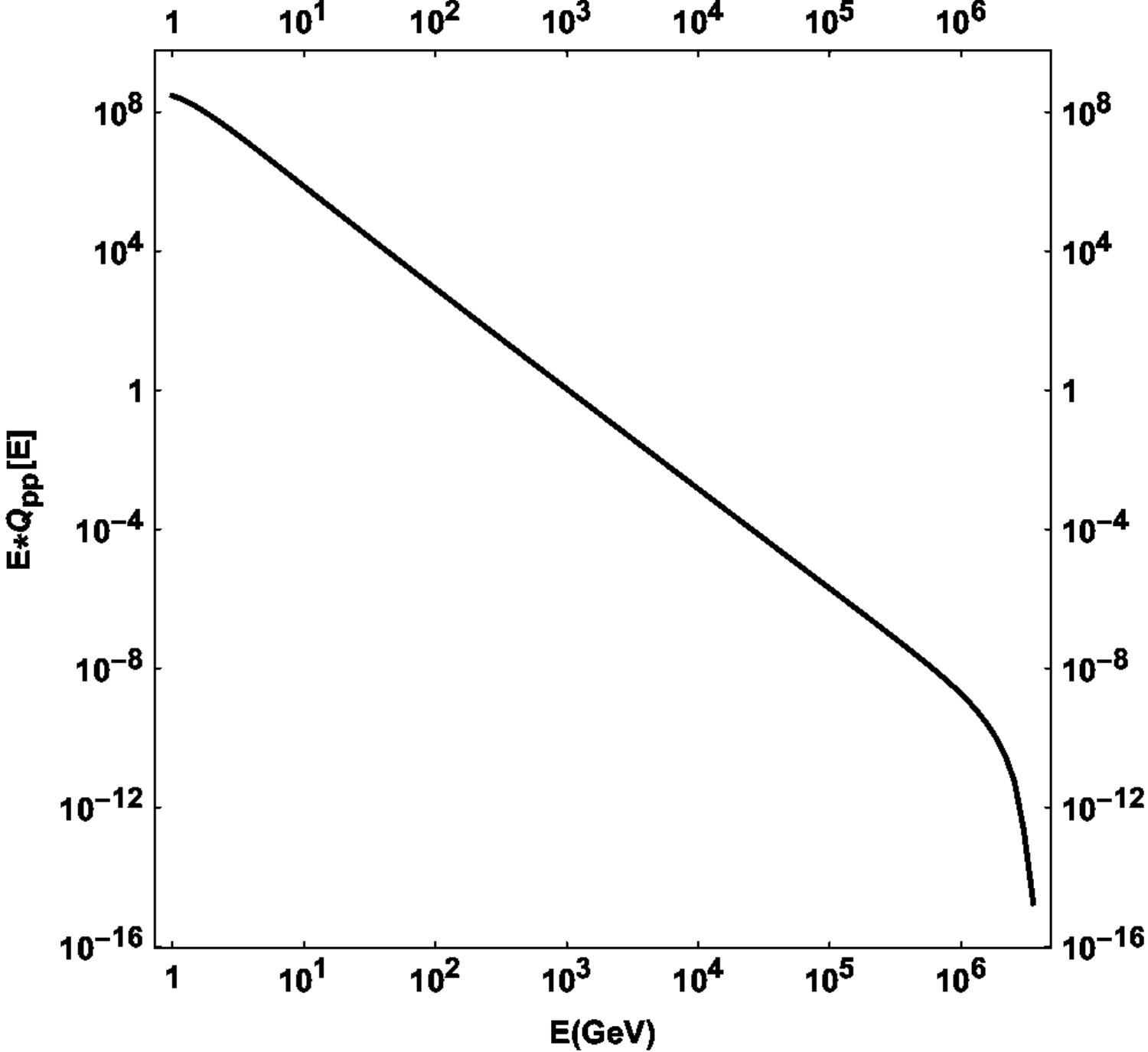}
\caption{Pion injection function Q, weighted by pion energy, measured in non-normalized units, describing the spectrum emerging from many (non-thermal--thermal) p--p collisions. Contributions rapidly decline as particle energy increases. This figure uses a test velocity vector of (0.2, 0.8, 0.1)c.} 
\label{qpp_plot}
\end{figure}
\begin{figure}[H]\vspace{-6pt}
\includegraphics[width=10.5 cm]{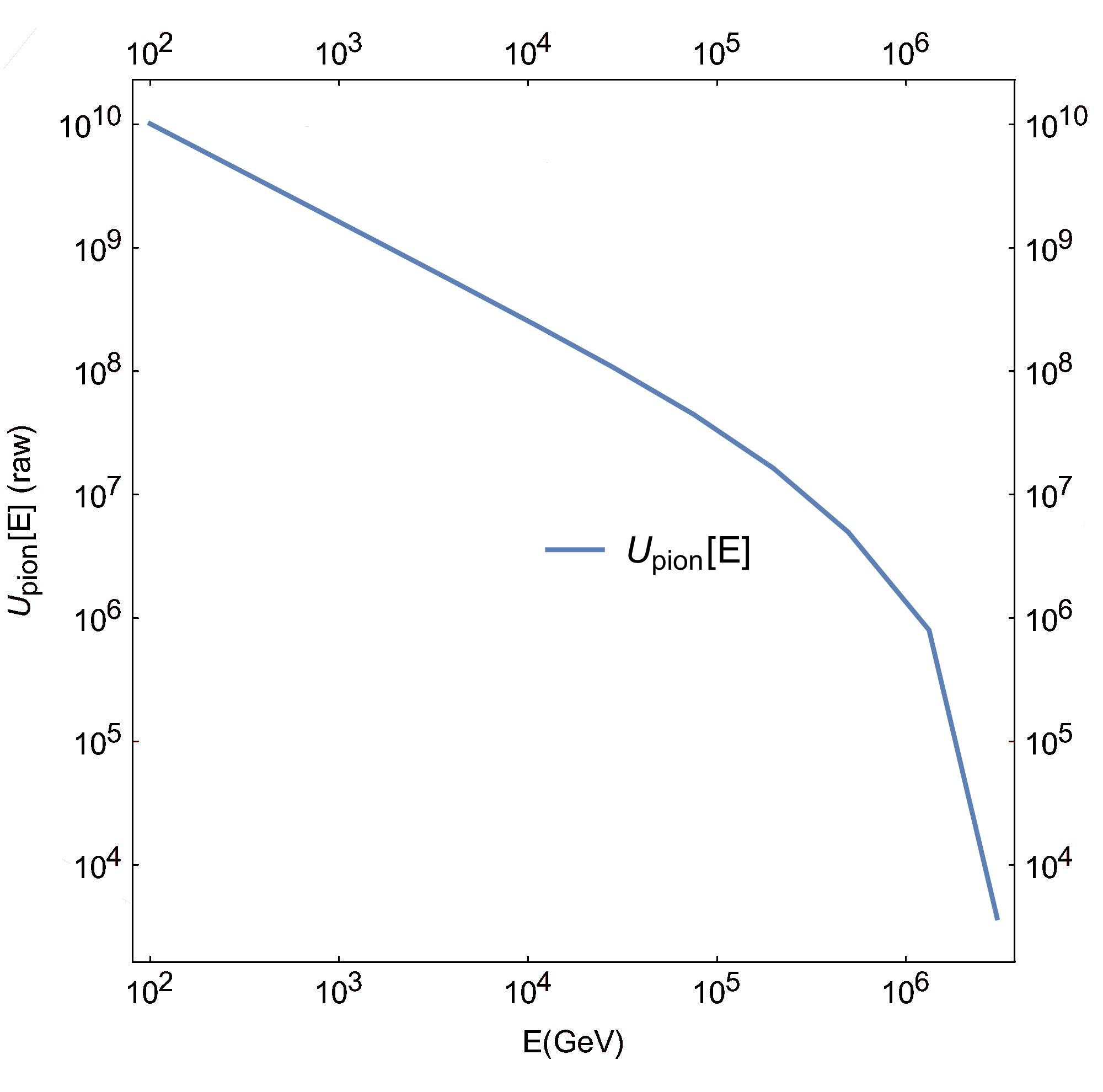}
\caption{Pion energy distribution plotted in non-normalized units versus energy. In the software, the above distribution is represented by  function U.}
\label{upp_plot}
\end{figure}

\textls[-15]{A number of energy loss mechanisms are included for the hot proton \mbox{distribution \cite{Kelner2006,Reynoso2009,Reynoso2019}}}. The discussion of this subsection is based on a cell with the following properties: ($u_{x}, u_{y},$ $u_{z}, b_{x}, b_{y}, b_{z}, n_{\mathrm{p}}, \phi_{1}, \phi_{2}, \alpha$) = ($-$0.3780c, 0.4480c, 0.0124c, 10$^{5}$ G, 10$^{6}$ G, 10$^{5}$ G, 2.1 $\times$ 10$^{11}$cm$^{-3}$, 1.047 rad, 5.00 $\times$ 10$^{-7}$~rad, 2.0), where u stands for velocity and b for magnetic field along directions x, y, or z. In the following, n is the jet proton density, and $\alpha$ is the high-energy proton distribution spectral index.  As an exception, Figure \ref{qpp_plot} uses a velocity vector of (0.2, 0.8, 0.1)c.
Figures \ref{sigma_inel_pp}--\ref{upp_plot} are taken from \cite{Smponias_2021}.

A hot proton upper energy cutoff at E $\le$ 10$^{6}$ GeV is used. The adiabatic system expansion time scale is taken to be \cite{Reynoso2009}
\begin{equation}
t_{\mathrm{adb}}^{-1}=\frac{2}{3} \frac{ u_{ \mathrm{b}(\mathrm{adb})} }{ z_{\mathrm{j}} }
\end{equation}
with z$_{\mathrm{j}}$ = 10$^{11}$ cm  being the characteristic width of the jet system at the neutrino emission size scale. For this calculation, $u_{\mathrm{b}(\mathrm{adb})}$ is preset to 0.8, for a 0.8c jet.

For the p--p collision energy loss mechanism,
\begin{equation}
t_{\mathrm{pp}}^{-1}=n c \sigma_{\mathrm{inel_{pp}}}(E_{\mathrm{p}}) K_{\mathrm{pp}}
\label{tpp_eqn1}
\end{equation}
where n is the bulk flow number density, K$_{\mathrm{pp}}$ = 0.5 \cite{Reynoso2009}. Equation (\ref{tpp_eqn1}) is justified if we consider a small cube of matter of number density n, moving at speed (near) c, having a surface A perpendicular to its direction of motion. Then, n*c has the dimensions of \mbox{cm$^{-2}$ $\times$ s$^{-1}$}. This is then multiplied by the inelastic p--p collision cross-section \cite{Reynoso2009} 
\begin{equation}
\sigma^{(\mathrm{inel})}_{\mathrm{pp}}=(34.3+1.88L+0.25L^{2}) \times [1-(\frac{E_{\mathrm{th}}}{E_{\mathrm{p}}})^{4}]^{2} \times 10^{-27} \mathrm{cm^{2}} 
\end{equation}
L = ln(E$_{\mathrm{p}}$/1000GeV), E$_{\mathrm{th}}$ = 1.2 GeV. In Figure \ref{sigma_inel_pp}, $\sigma^{(\mathrm{inel})}_{\mathrm{pp}}$ is plotted. 

The pion decay timescale is 
\begin{equation}
t_{\pi}=t_{\pi 0} \Gamma _{\pi} +t_{\mathrm{esc}}
\end{equation}
where $\Gamma_{\pi}$ is the pion Lorentz factor and 
\begin{equation}
t_{\pi 0}=2.6 \times 10^{-8} \mathrm{s}
\end{equation}

In the calculations, we employ the form
\begin{equation}
t_{\pi}=t_{\pi 0} (\frac{E_{\pi}}{m_{\pi} c^{2}}) +t_{\mathrm{esc}}
\end{equation}
where m$_{\pi}$ is the pion mass. The light escape time $t_{\mathrm{esc}}$ is a sensitive parameter of the model.

The synchrotron hot proton energy loss time scale is \cite{Reynoso2009}
\begin{equation}
t_{\mathrm{sync}}^{-1}=\frac{4}{3} \left(\frac{ m_{\mathrm{e}} }{ m_{\mathrm{p} }  } \right)^{3} \frac{1}{8 \pi c m_{\mathrm{e}}} \sigma_{\mathrm{T}} B^{2} \frac{E_{\mathrm{p}}}{m_{\mathrm{p}} c^{2}}
\end{equation}


m$_{\mathrm{e}}$ is the electron mass and m$_{\mathrm{p}}$ the proton mass. $\sigma_{T}$=$\frac{8 \pi}{3} (\frac{e^{2}}{m_{\mathrm{e}} c^{2}})^{2}$ = 6.65 $\times$ 10$^{-25}$ cm$^{2}$ is the Thompson cross-section, and B is the local magnetic field. The proton's Lorentz factor $\Gamma_{\mathrm{p}}$ is written as $\frac{E_{\mathrm{p}}}{m_{\mathrm{p}} c^{2}}$, in order to facilitate energy-dependent calculations later on. In~summary,
\begin{equation}
t_{\mathrm{loss}}^{-1}=t_{\mathrm{sync}}^{-1}+t_{\mathrm{adb}}^{-1}+t_{\mathrm{pp}}^{-1}
\end{equation}

The time-scales of the different energy-loss mechanisms are presented in Figure \ref{tloss_many}.


\subsection{Particle Cascades in the Jets} 

Hot--cold proton interaction results to a distribution of high-energy pions, which then decay, allowing for the creation of energetic neutrinos. We have \cite{KS18,SK17,SK15,Campion2020}
\begin{equation}
pp \rightarrow pp \pi^{0}+\pi_{0} \, ,
\end{equation}
for neutral pions $\pi^{0}$, and
\begin{equation}
pp \rightarrow pn \pi^{+} \, , \qquad pp \rightarrow pn \pi^{-}+  \pi^{+}+\pi^{+} \, ,
\end{equation}
for $\pi^{\pm}$.

Neutral pions $\pi^0$ decay to gamma rays. On the other hand, $\pi^{\pm}$ mainly decay to an antimuon and a muonic neutrino, or to a muon and an antineutrino (prompt neutrinos)~\mbox{\cite{KS18,Campion2020}}. We therefore obtain neutrinos that escape the system, after the cascade.
\begin{equation}
\pi^{+} \rightarrow	\mu^{+} + \nu_{\mu} \, ,  \qquad 
\pi^{-} \rightarrow \mu^{-} + \widetilde{\nu}_{\mu} \, .
\label{muon-produc}
\end{equation}
{As} an approximation, we do not consider neutrino production through secondary channels or delayed neutrinos.

For each successive particle population in the above cascades, the transport equation for nonstochastic phenomena and for time-independent transport (transport time much less than the time step of the dynamic simulation), takes a simplified form:
\begin{equation}
\frac{\partial N}{\partial E} + \frac{N}{t_{\mathrm{loss}}} = Q(E,\vec{r})
\end{equation}
where $\vec{r}$ is the spatial position vector, N is the particle density of the daughter  population, and Q is its injection function. 

A power-law distribution is assumed for protons (Figure \ref{fast-p-density} shows power-law fast proton density), replacing the solution of the first transport equation in the cascade. Afterwards, along the cascade, we calculate the properties of resulting particle distributions~\cite{Kelner2006}. 


\subsection{Lorentz Transform of High E Proton Distribution}

For the calculation of the fast proton distribution, the relevant directional equation (direction is defined by the angle $\theta$ between  velocity and line of sight) can be found \mbox{in \cite{TR11,PS2001}}. 
The selected variant originates from \cite{TR11}, minus a geometry factor that is here absorbed into the normalization factor \cite{Smponias_2021}
\begin{equation}
n(E,\theta)=\frac{\Gamma^{- \alpha -1} E^{- \alpha} (1- \beta \mathrm{cos}(\theta) \sqrt{1- \frac{m^{2} c^{4}}{E^{2}} } )^{- \alpha -1} }{[\mathrm{sin^{2}}(\theta) + \Gamma^{2}(\mathrm{cos}(\theta) - \frac{\beta}{\sqrt{1- \frac{m^{2} c^{4} }{E^{2}}}})^{2}]^{\frac{1}{2}}  }
\label{tr13} 
\end{equation}
where $\Gamma$ is the Lorentz factor of the particles in a tiny volume.




\subsection{Pion Injection Function and Pion Energy Distribution}


For each fast--slow proton interaction, a spectrum of possible pion energies exists, given by  function $F_{\pi}$ \cite{Kelner2006,Reynoso2009,Reynoso2019}. \vspace{-12pt}
\begin{adjustwidth}{-\extralength}{0cm}
\centering 
\begin{eqnarray}
F_{\pi}^{(\mathrm{pp})}\left( x,\frac{E}{x} \right) = 4 \alpha B_{\pi} x^{\alpha-1} 
\left( \frac{1-x^{\alpha}}{1+r x^{\alpha}(1-x^{\alpha})} \right)^{4}  
\left( \frac{1}{1-x^{\alpha}} + \frac{r(1-2x^{\alpha})}{1+rx^{\alpha}(1-x^{\alpha})} 
\right) \left( 1- \frac{m_{\pi} c^{2}}{x E_{\mathrm{p}}}  \right)^{\frac{1}{2}} 
\label{Ffunction}
\end{eqnarray}
\end{adjustwidth}
where 
 $x=E/E_{\mathrm{p}}$, B$_{\pi}$ = $\alpha^{'}$ + 0.25, $\alpha^{'}$ = 3.67 +0.83L + 0.075L$^{2}$,  r = 2.6/$\sqrt{\alpha^{'}}$, $\alpha$ = 0.98/$\sqrt{\alpha^{'}}$
 \cite{Reynoso2009,Kelner2006}. 

 
Figure \ref{Fpp_plot} $xF_{\pi}$ is plotted with the fraction $x$ for different fast proton energies.

Pion injection function $Q_{\pi}^{(\mathrm{pp})}$ comprises pion contributions at each pion energy to that pion energy from  spectrum F of all potential p--p interactions. 
\begin{eqnarray}
Q_{\pi}^{(\mathrm{pp})}(E,z) = n(z) c \int \limits_{\frac{E}{E_{\mathrm{p}}^{(\mathrm{max})}}}^{1} \frac{dx}{x}
\left(  \frac{E}{x},z \right) F_{\pi}^{(\mathrm{pp})} \left( x,\frac{E}{x} \right) 
\sigma^{(\mathrm{inel})}_{\mathrm{pp}} \left( \frac{E}{x} \right)  \, ,
\label{Qpp}
\end{eqnarray}
$x$ is the fraction of the pion energy to proton energy, and $n(z)$ is the jet flow proton density. 

 Figure \ref{qpp_plot} plots $Q_{\pi}^{(\mathrm{pp})}$  versus  pion 
energy $E_\pi$.


In order to obtain the pion distribution, the following transport equation is employed:  
\begin{equation}
\frac{\partial N_{\pi}}{\partial E} + \frac{N_{\pi}}{t_{\mathrm{loss}}} = Q_{\pi}^{(\mathrm{pp})}(E,z)
\end{equation}
where N$_{\pi}(E,z)$ denotes the pion energy distribution. Then
\begin{eqnarray}
N_{\pi}(E) = \frac{1}{|b_{\pi}(E)|} \int \limits_{E}^{E^{(\mathrm{max})}} dE' Q_{\pi}^{(\mathrm{pp})}(E') 
\exp {[-\tau_{\pi}(E,E')]} \, ,
\end{eqnarray}
and 
\begin{eqnarray}
\tau_{\pi}(E',E)=\int \limits_{E'}^{E} \frac{dE'' t_{\pi}^{-1}(E)}{|b_{\pi}(E'')|} \, .
\end{eqnarray}
$b_{\pi (E)}=-E(t_{\mathrm{sync}}^{-1}+t_{\mathrm{adb}}^{-1}+t^{-1}_{\pi p}+t^{-1}_{\pi \gamma}) $ is the energy loss rate of the pion. As an approximation, the last term in the latter expression is omitted. In Figure \ref{upp_plot} plots nemiss \cite{nemiss} software function U (U$_{\mathrm{analytical}}$), representing N$_{\pi}$, with pion energy.

The above calculations are performed for each computational cell. A cell is macroscopically large inasmuch as only the deterministic portion of the transport equation is employed, in turn rendering it deterministic. Again, we take the characteristic scale (mean free path) of the radiative interactions to be smaller than the cell size, leading to the containment of particle interactions within a given hydrocode cell. Furthermore, the time scale for the radiative interactions is taken to be smaller enough than the hydrocode's time step, so that a cell's radiative interactions belong to a single time step each time.    


\subsection{Neutrino Emissivity}

As mentioned in the main text, the emissivity of prompt neutrinos \cite{Kelner2006,Lipari2007,Reynoso2008,Reynoso2009}, is 
\begin{eqnarray}
Q_{\pi \rightarrow \nu}(E) = \int \limits_{E}^{E_{\mathrm{max}}} dE_{\pi} t^{-1}_{\pi} 
(E_{\pi}) N_{\pi}(E_{\pi}) \frac{\Theta (1-r_{\pi}-x)} {E_{\pi}(1-r_{\pi})}  \, ,
\label{Neut-Emiss}
\end{eqnarray}
E is neutrino energy, $x=E/E_{\pi}$, r$_{\pi}$=(m$_{\mu}$/m$_{\pi}$)$^{2}$ and $t_{\pi}$ is the pion decay timescale. $\Theta$($\chi$) 
is the theta function ~\cite{Reynoso2009,SK15}. 
Neutrino emissivity is calculated for each individual cell using the angle of the local velocity to the LOS crossing that cell.


\section{Flow Diagrams}\label{appe}

In this section, some code flow diagrams are provided, as supplemental material.

\begin{figure}[H]

\begin{adjustwidth}{-\extralength}{0cm}
\centering 
\includegraphics[width=18.2 cm]{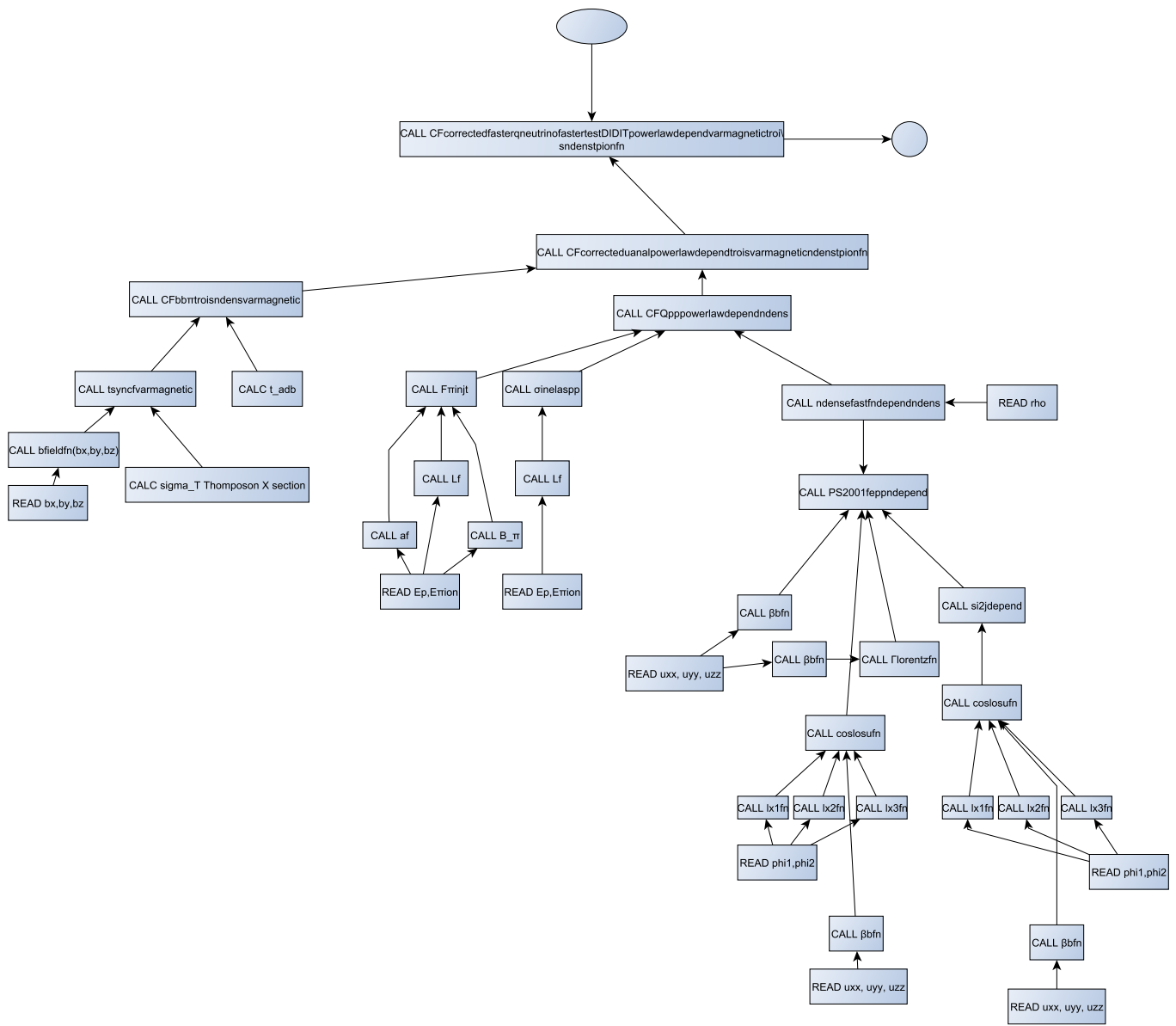}
\end{adjustwidth}
\caption{Flow diagram of nemiss depicting the basic structure of the programme.}
\label{flow_nemiss}
\end{figure}

\begin{figure}[H]
\includegraphics[width=12 cm]{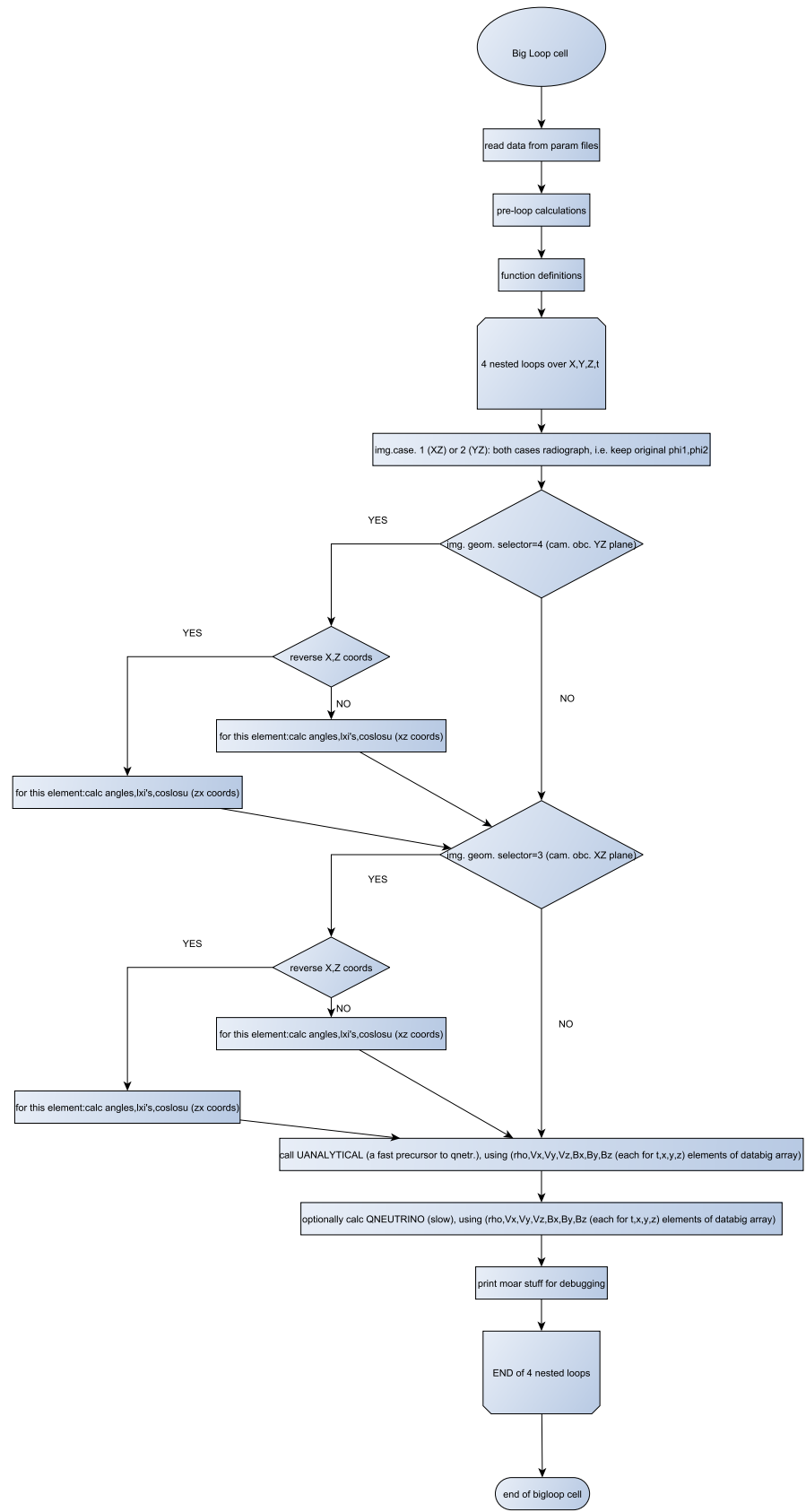}
\caption{Flow diagram of the main loop of nemiss where the calculation takes place. This was upgraded to 5D, with the addition of a loop for particle energy.}
\label{flow_nemiss_loop}
\end{figure}

\begin{figure}[H]

\begin{adjustwidth}{-\extralength}{0cm}
\centering 
\includegraphics[width=17 cm]{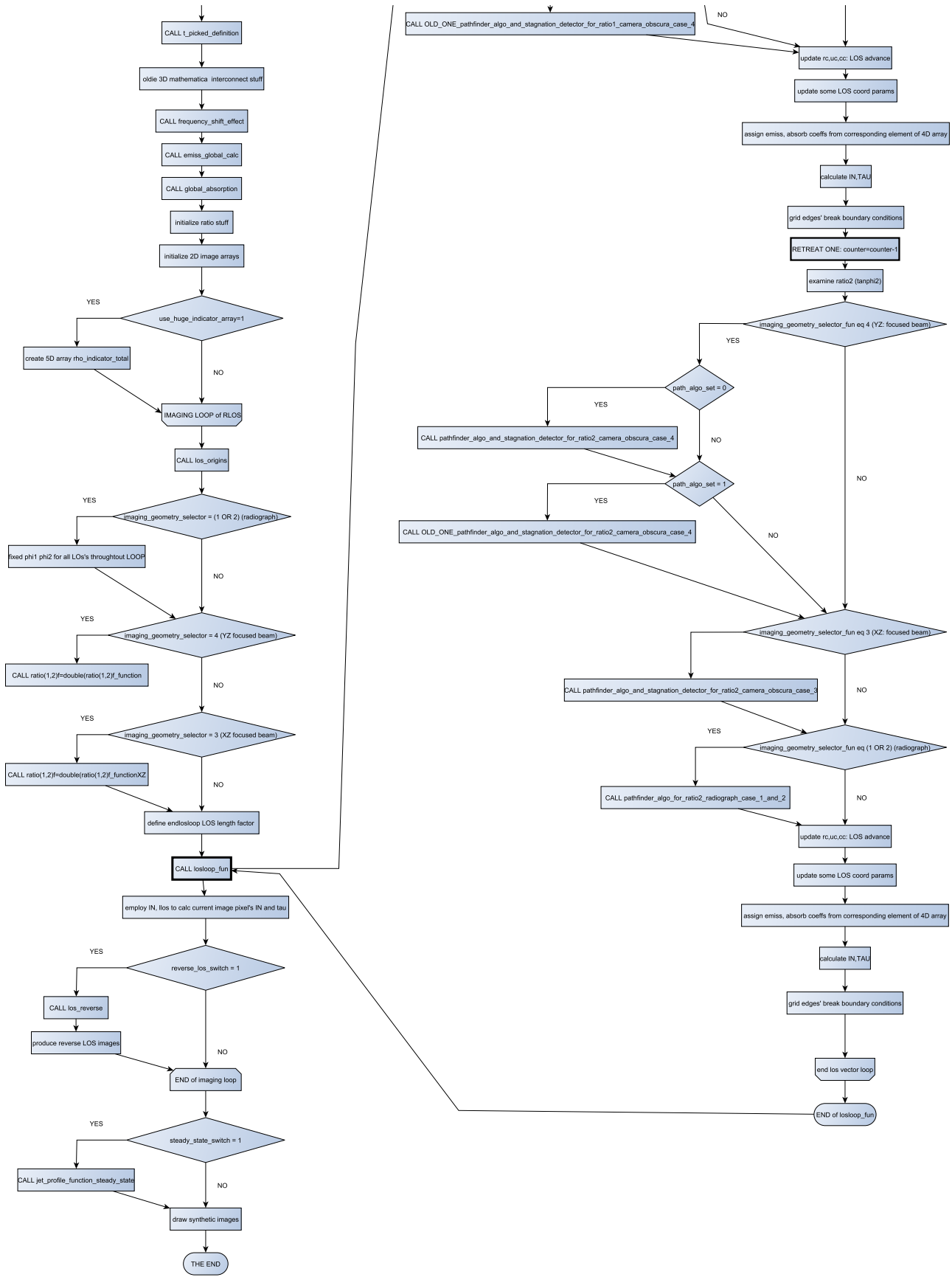}
\end{adjustwidth}
\caption{{Flow diagram of rlos.1 depicting} 
 procedures and functions called during programme execution.}
\label{flow_rlos}
\end{figure}



\begin{adjustwidth}{-\extralength}{0cm}

\reftitle{References}



\PublishersNote{}
\end{adjustwidth}

\end{document}